\documentclass[twocolumn]{aastex631}

\received{\today}
\revised{\today}
\accepted{\today}
\submitjournal{ApJ}

\usepackage{amsmath}	
\usepackage{amssymb}
\usepackage[abs]{overpic}
\usepackage{xcolor,varwidth}
\usepackage[normalem]{ulem}

\newcommand{\rg}{r_{\rm g}}
\newcommand{\tg}{r_{\rm g}/c}
\newcommand{\Jtot}{\dot{J}_{\rm tot}}
\newcommand{\Mdot}{\dot{M}}
\newcommand{\Jadv}{\dot{J}_{\rm adv}}
\newcommand{\Jout}{\dot{J}_{\rm stress}}
\newcommand{\JoutM}{\dot{J}_{\rm stress, M}}
\newcommand{\JoutR}{\dot{J}_{\rm stress, R}}

\newcommand{\alphaM}{\alpha_{\rm M}}
\newcommand{\alphaR}{\alpha_{\rm R}}
\newcommand{\hamr}{{H-AMR}}

\shorttitle{Angular momentum transport in MADs}
\shortauthors{Chatterjee \& Narayan}

\graphicspath{{./}{figures/}}
\begin{document}

\title{Flux eruption events drive angular momentum transport in magnetically arrested accretion flows}

\correspondingauthor{Koushik Chatterjee}
\email{koushik.chatterjee@cfa.harvard.edu}

\author[0000-0002-2825-3590]{K. Chatterjee}\affiliation{Black Hole Initiative at Harvard University, 20 Garden Street, Cambridge, MA 02138, USA}\affiliation{Harvard-Smithsonian Center for Astrophysics, 60 Garden Street, Cambridge, MA 02138, USA}

\author{R. Narayan}\affiliation{Black Hole Initiative at Harvard University, 20 Garden Street, Cambridge, MA 02138, USA}\affiliation{Harvard-Smithsonian Center for Astrophysics, 60 Garden Street, Cambridge, MA 02138, USA}

\begin{abstract}

\noindent We evolve two high-resolution general relativistic magnetohydrodynamic (GRMHD) simulations of advection-dominated accretion flows around non-spinning black holes (BHs), each over a duration $\sim 3\times 10^5\,GM_{\rm BH}/c^3$. One model captures the evolution of a weakly magnetized (SANE) disk and the other a magnetically arrested disk (MAD). Magnetic flux eruptions in the MAD model push out gas from the disk and launch strong winds with outflow efficiencies at times reaching $10\%$ of the incoming accretion power. Despite the substantial power in these winds, average mass outflow rates remain small out to a radius $\sim100\,GM_{\rm BH}/c^2$, only reaching $\sim 60-80\%$ of the horizon accretion rate. The average outward angular momentum transport is primarily radial in both modes of accretion, but with a clear distinction: magnetic flux eruption-driven disk winds cause a strong vertical flow of angular momentum in the MAD model, while for the SANE model, the magnetorotational instability (MRI) moves angular momentum mostly equatorially through the disk. Further, we find that the MAD state is highly transitory and non-axisymmetric, with the accretion mode often changing to a SANE-like state following an eruption before reattaining magnetic flux saturation with time. The Reynolds stress changes direction during such transitions, with the MAD (SANE) state showing an inward (outward) stress, possibly pointing to intermittent MRI-driven accretion in MADs. Pinning down the nature of flux eruptions using next-generation telescopes will be crucial in understanding the flow of mass, magnetic flux and angular momentum in sub-Eddington accreting BHs like M87$^*$ and Sagittarius A$^*$.

\end{abstract}


\section{Introduction} \label{sec:intro}

\noindent The 2017 Event Horizon Telescope (EHT) Collaboration results on the supermassive black holes (BHs), Sagittarius A$^*$ (or Sgr A$^*$) and M87$^*$, suggest that these BHs are fed by gas with dynamically-important magnetic fields \citep[][]{EHT_M87_2019_PaperV,EHT_M87_2019_PaperVIII,EHT_SgrA_2022_PaperV} that can potentially affect the evolution of the BH's environment. Further, we know that these BHs accrete at highly sub-Eddington rates in the form of a hot, two-temperture, advection-dominated accretion flow \citep[ADAF,][]{nar94,nar95a,abr95,Shapiro:1976,Ichimaru:1977,ree82,Yuan_2014}. Such systems are known to have low luminosities relative to their accretion rates \citep[e.g.,][]{Narayan:95a,Yuan:03,Bower:03,Marrone:07,Kuo:2014}. It is still unknown how magnetic fields determine the evolution of mass and angular momentum in ADAFs. A few numerical simulations have attempted to disentangle the highly non-linear coupling of magnetic fields, gas and extreme gravity to understand mass loss via disk turbulence and wind/jet outflows \citep[e.g.,][]{penna10,Narayan:2012,Yuan:12_winds,Sadowski:13,White:2020_RIAF,Ressler:2020:sgra_MAD,Begelman2022}. But much remains to be understood.

\begin{figure*}
    \includegraphics[width=0.5\textwidth]{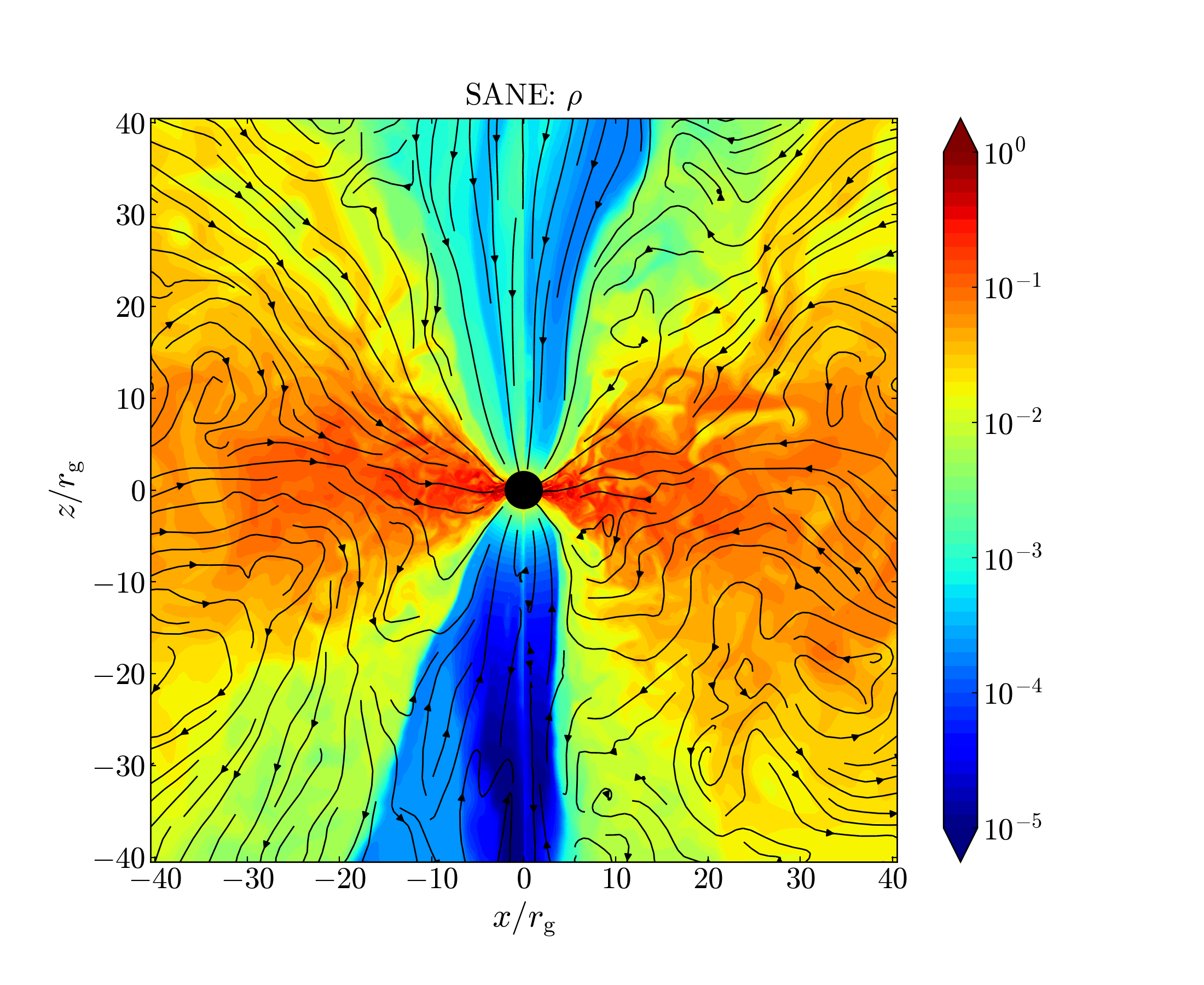}
    \includegraphics[width=0.5\textwidth]{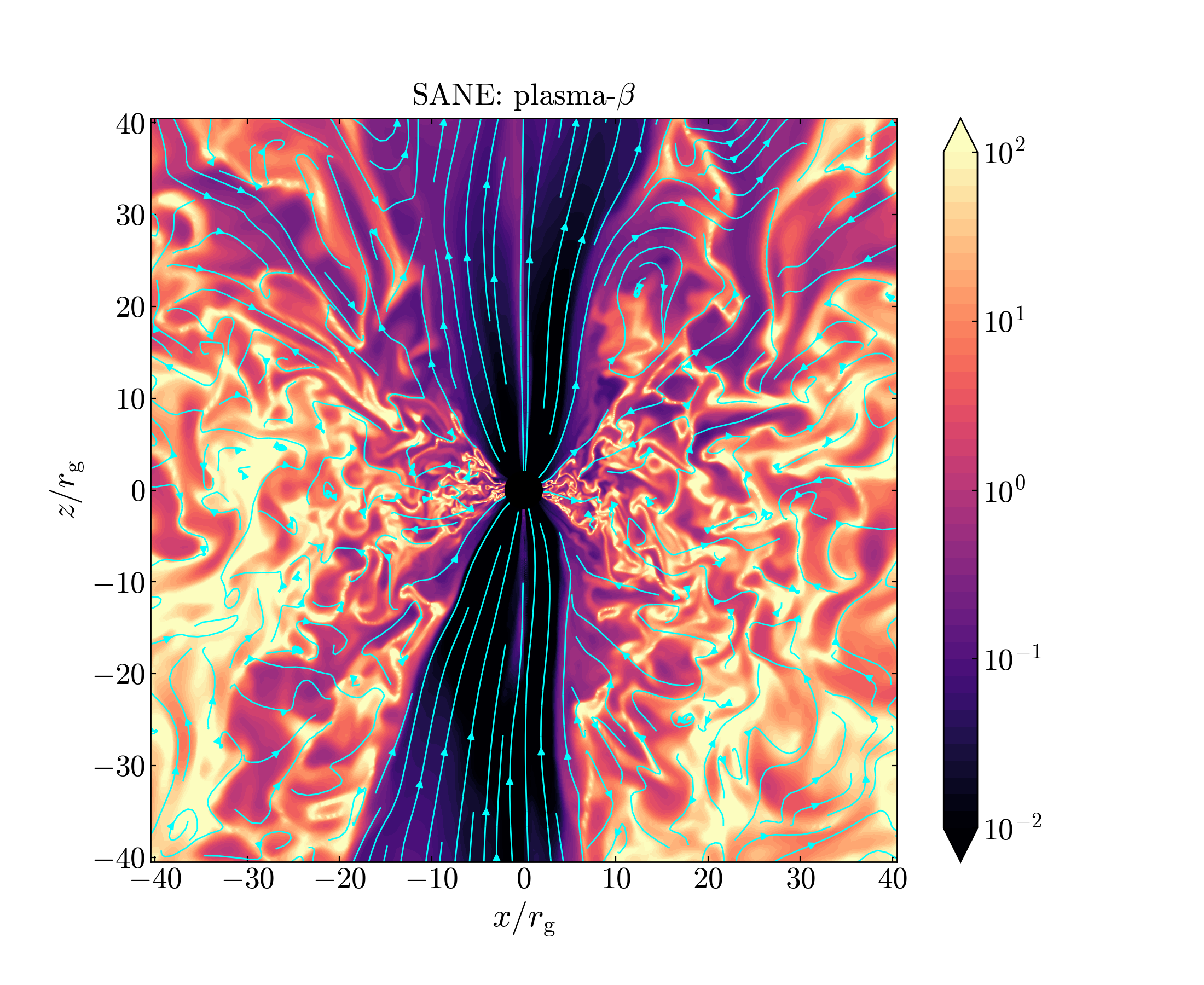}
    \includegraphics[width=0.5\textwidth]{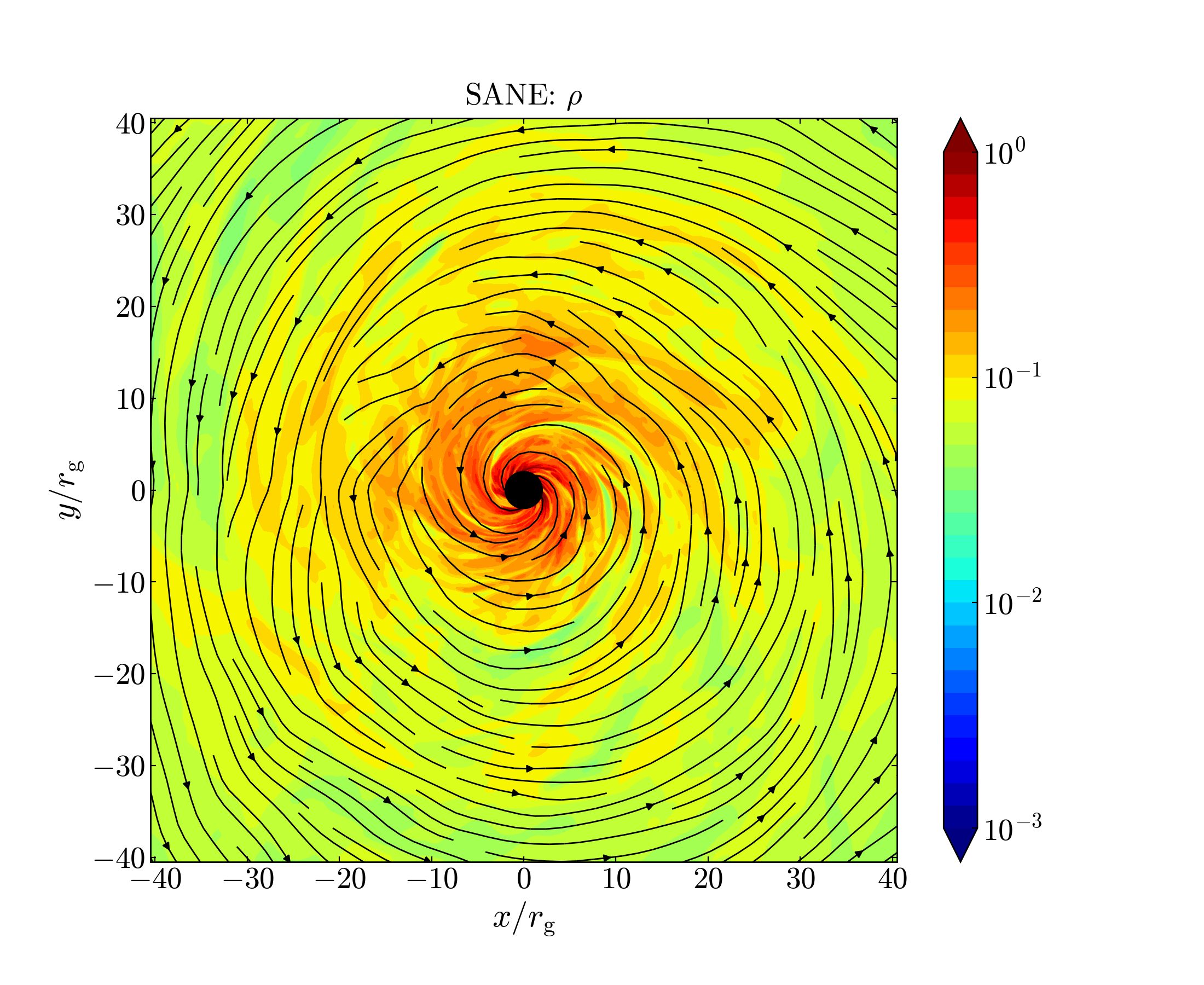}
    \includegraphics[width=0.5\textwidth]{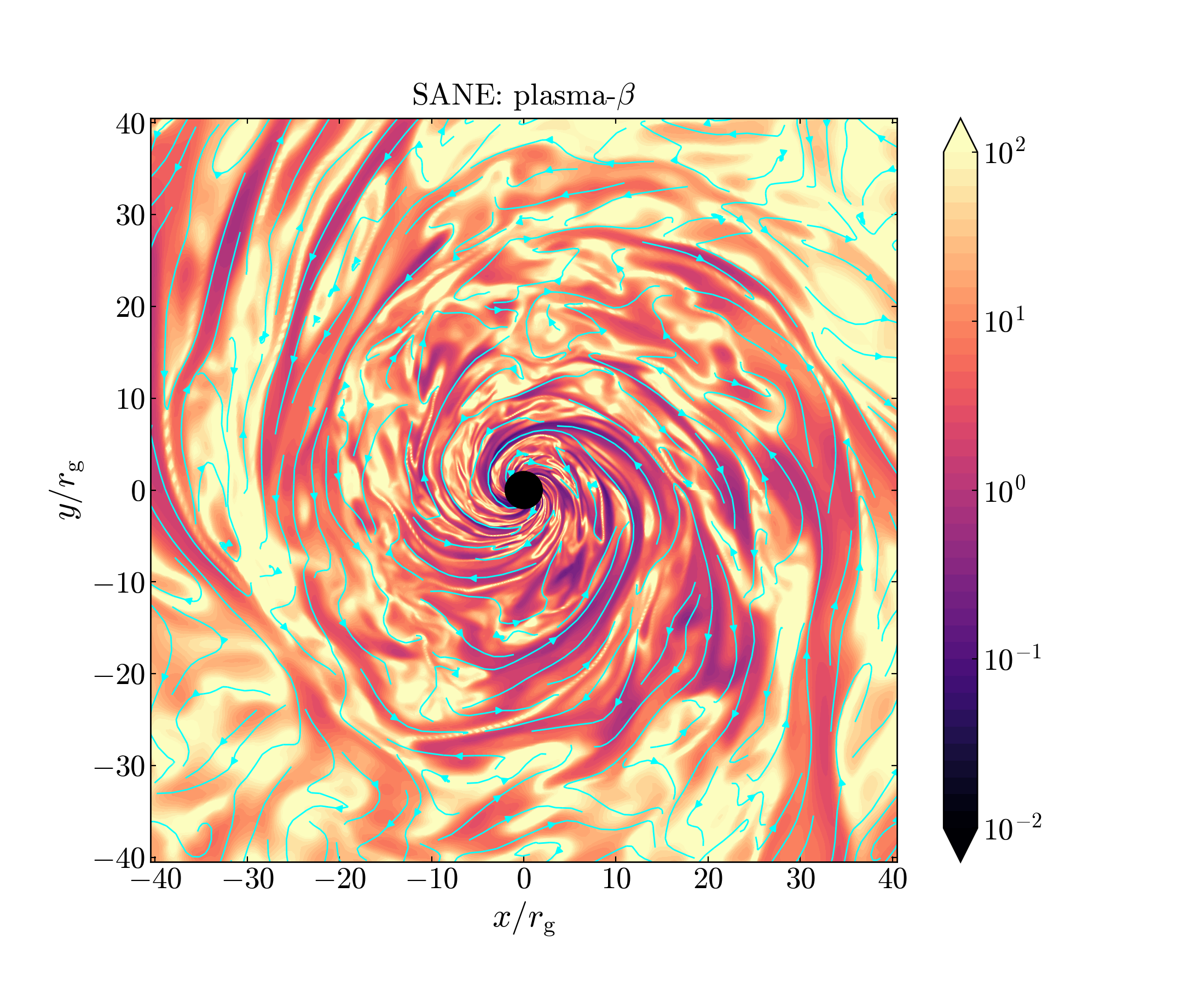}
	\caption{We show a snapshot of the weakly magnetized (SANE) simulation at $t=1.3\times10^5\tg$. Top row shows a vertical slice of the gas density and the plasma-$\beta\,(\equiv p_{\rm gas}/p_{\rm mag})$, while the bottom row shows the midplane cross-section. The black lines correspond to the velocity streamlines and the cyan lines denote the magnetic field lines. The disk exhibits laminar gas inflow in the midplane punctuated by small-scale turbulent eddies as seen from the plasma-beta.}
    \label{fig:SANE_slice}
\end{figure*}

\begin{figure*}
    \includegraphics[width=0.5\textwidth]{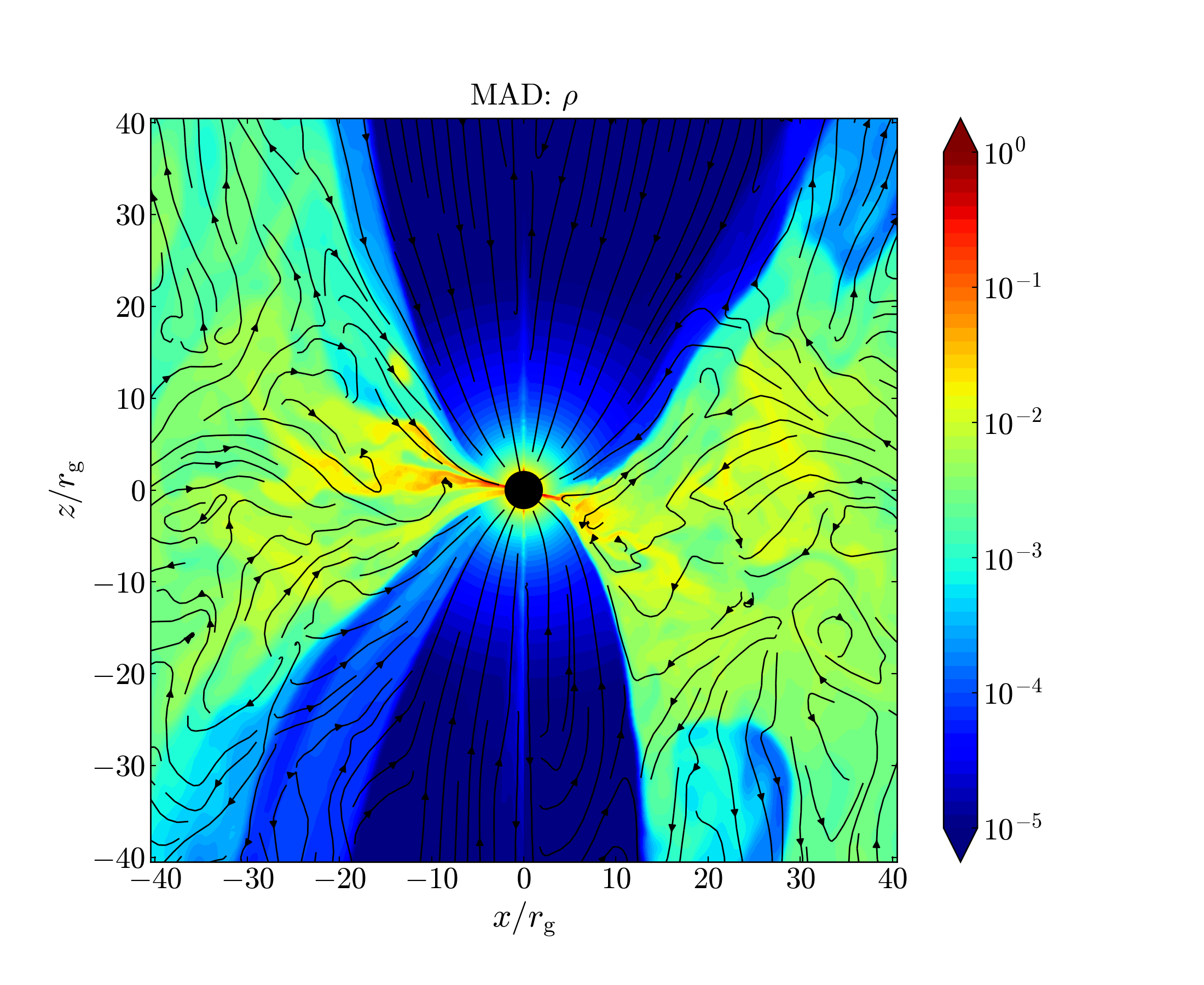}
    \includegraphics[width=0.5\textwidth]{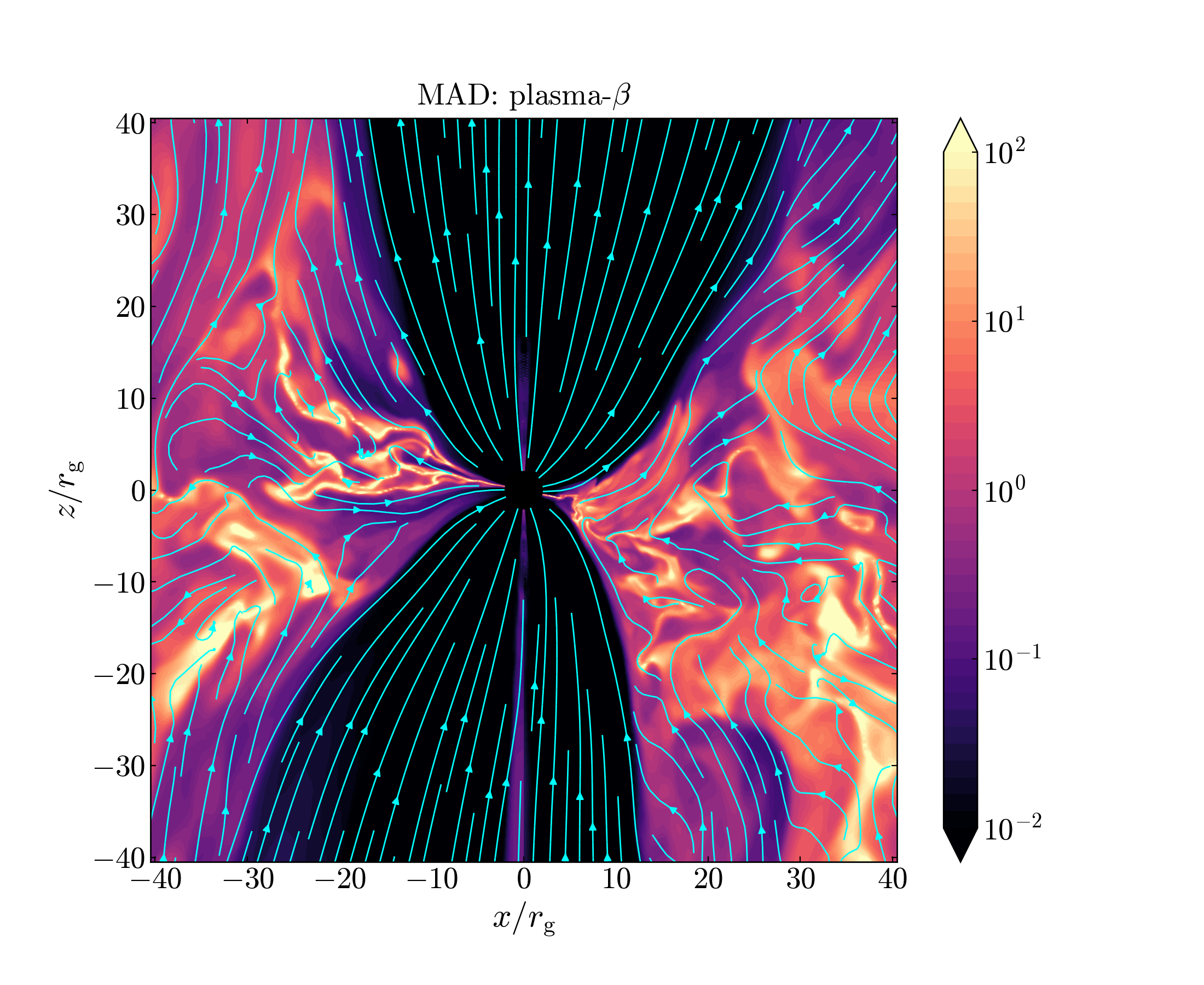}
    \includegraphics[width=0.5\textwidth]{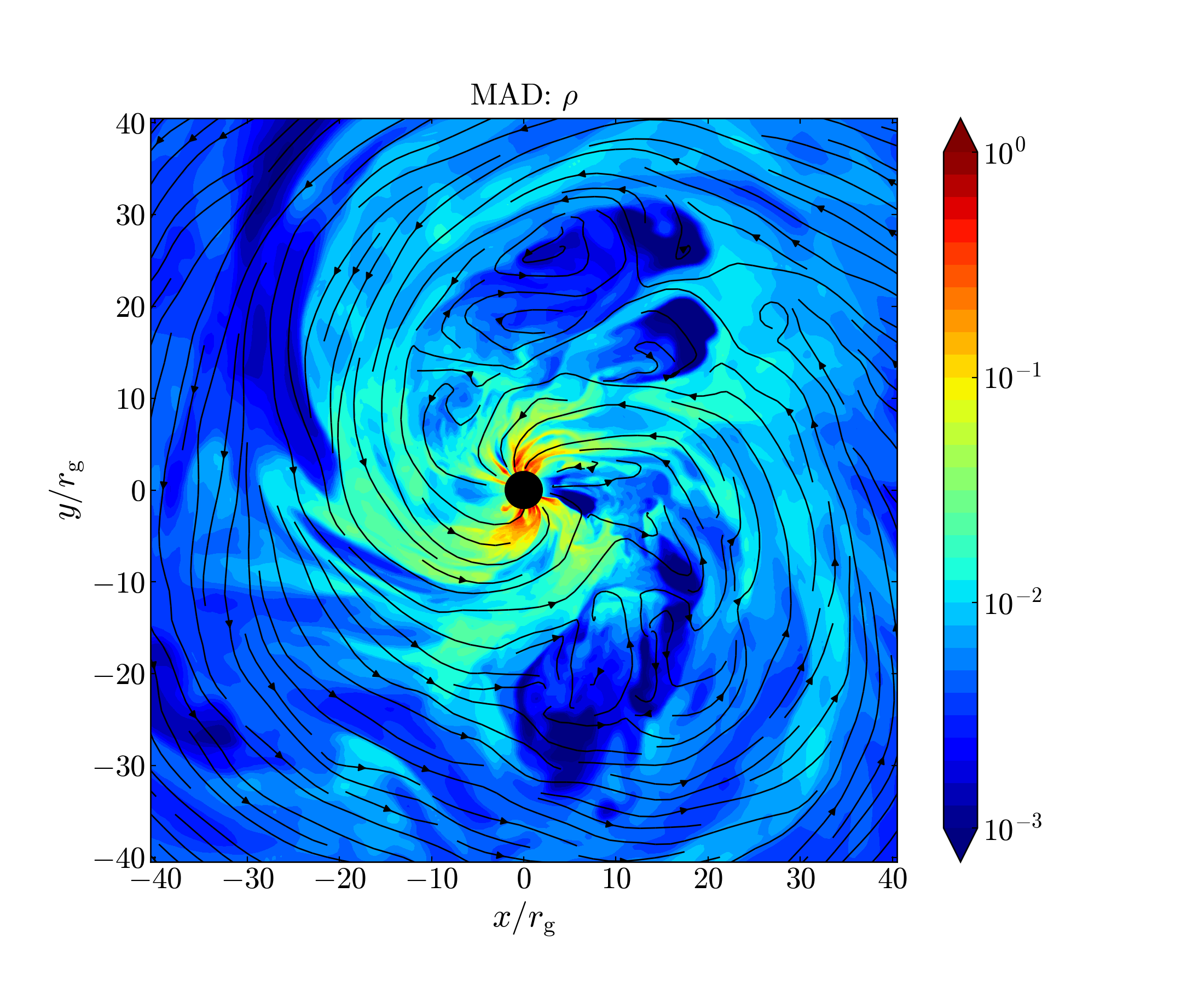}
    \includegraphics[width=0.5\textwidth]{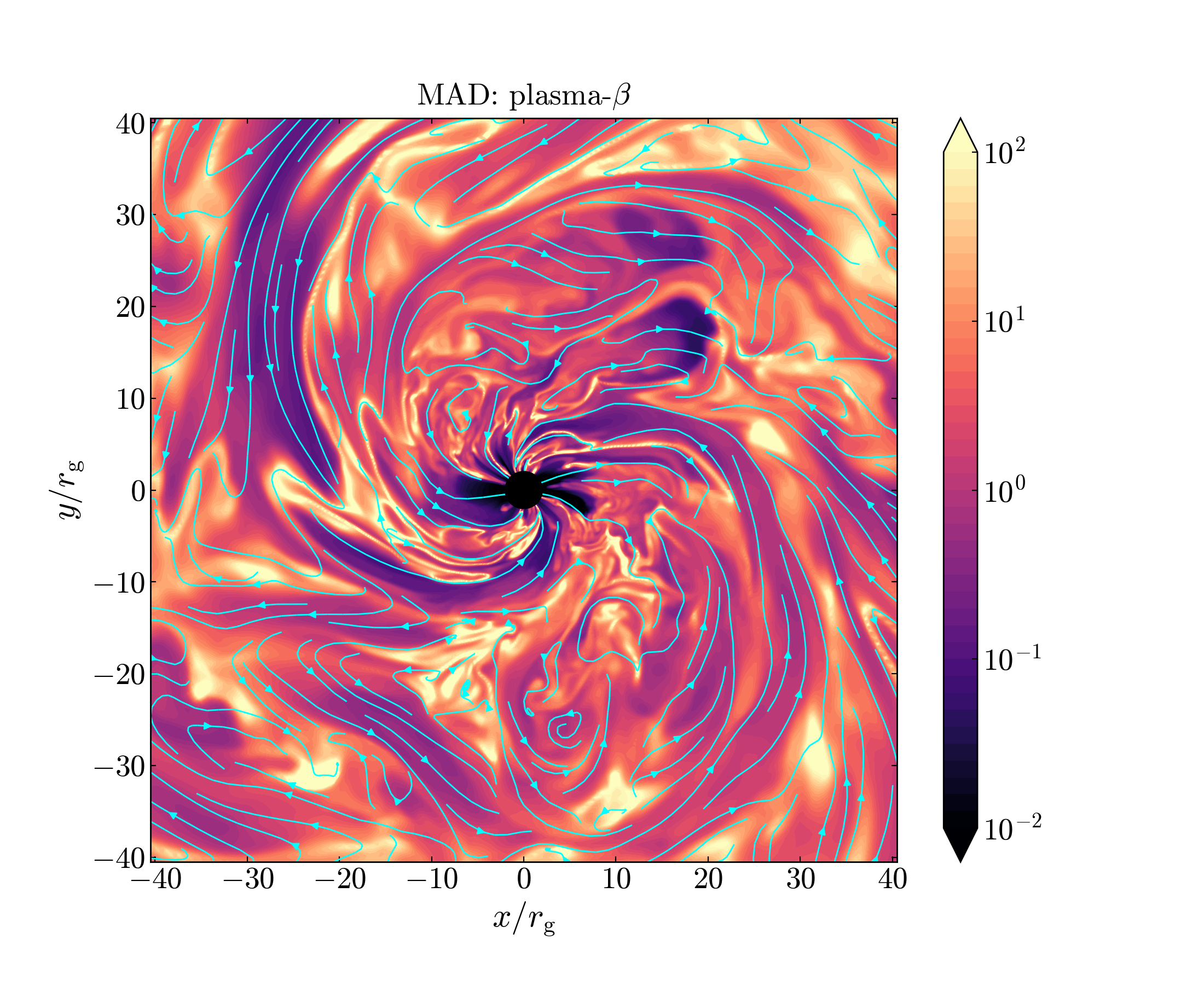}
	\caption{Same as Fig.~\ref{fig:SANE_slice}, but for the magnetically arrested disk (MAD) simulation at $t=2.9\times10^5\tg$. In contrast to the SANE model, the inflow is broken up by outgoing magnetic flux-tubes and accretion occurs via interchange instabilities. We also see a wider polar vacuum region and a vertically thinner accretion flow near the black hole due to the presence of strong vertical magnetic fields.}
    \label{fig:MAD_slice}
\end{figure*}

Over the previous two decades, general relativistic magnetohydrodynamic (GRMHD) simulations have become a popular tool to study black hole accretion in various regimes, from sub-to-super Eddington accretion rates \citep[e.g.,][]{gam03,dev03code,mckinney06,tch11,Avara:2016,Sadowski:2016,Curd:2019,Porth:19,Liska:2022}. Generally, numerical simulations of ADAFs assume that an equilibrium hydrodynamic torus of gas \citep[e.g.,][]{fis76} feeds the BH with the help of the magneto-rotational instability \citep[MRI;][]{bal91}, which removes the disk's angular momentum to enable steady accretion. It is thought that the MRI is the main driver of accretion turbulence and perhaps, along with BH spin, leaves an indelible mark on horizon-scale observations by the EHT. Variability in the horizon-scale image and the multi-wavelength emission can be due to different causes, such as alternate accretion geometries, particle acceleration and radiative effects \citep[e.g.,][]{Chael_2019,Chatterjee_2020, Yoon:2020, Ressler:2020:sgra_MAD,Chatterjee:2021, Liska:2022, Lalakos:2022}. Indeed, while the parameter space of accretion models is vast, the near-horizon accretion structure is usually either near-Keplerian inspiralling gas or sub-Keplerian with dominant magnetic fields.

When enough magnetic flux is available in the disk, either by advecting magnetic fields from larger scales \citep[e.g.,][]{narayan03,Ressler:2020:sgra_MAD} or created in situ in the disk via dynamo mechanisms \citep[][]{liska_tor_2019}, the vertical fields near the BH can become strong enough to impede the accreting gas \citep[e.g.,][]{igu03,narayan03}. In such cases, MRI is thought to be suppressed due to magnetic pressure dominance in the disk. Accretion then proceeds primarily via magnetic interchange instabilities. Recently, however, \citet{Begelman2022} noted that it is actually the toroidal magnetic fields that dominate over the vertical fields during such accretion modes. Further the authors claim that the MRI is not completely suppressed but plays a major role in supporting the toroidal field and transporting angular momentum. Thus, understanding how the magnetic flux evolves in the disk is important in the study of angular momentum transport in MADs.

In this work, we revisit the standard torus models of both weakly and strongly magnetized accretion flows, simulated at high resolutions and over long timescales ($t\sim 3\times 10^5 GM_{\rm BH}/c^3$). Our interest is in the effect of strong magnetic fields and disk outflows on angular momentum transport. We focus on Schwarzschild (non-spinning) BHs so as to remove any influence from relativistic jets which could be powered by frame-dragging. Jets can remove a majority of angular momentum in the near-BH region \citep[e.g.,][]{tch12proc, Narayan2022} as well as initiate wind-jet mixing. These effects can lead to structured outflows and thus, enhance mass loss from the disk \citep[e.g., ][]{chatterjee2019}. We minimize such effects by restricting ourselves to non-spinning BHs. 

We describe our numerical setup in Sec.~\ref{sec:code}, and discuss the temporal and radial evolution of the disks in Sec.~\ref{sec:results}. We analyze the time-averaged and time-dependent angular momentum transport in Secs.~\ref{sec:jtot} and \ref{sec:MADtime}. We discuss astrophysical implications of our models in Sec.~\ref{sec:discussion} and conclude in Sec.~\ref{sec:conclusion}.

\begin{figure}
    \includegraphics[width=\columnwidth]{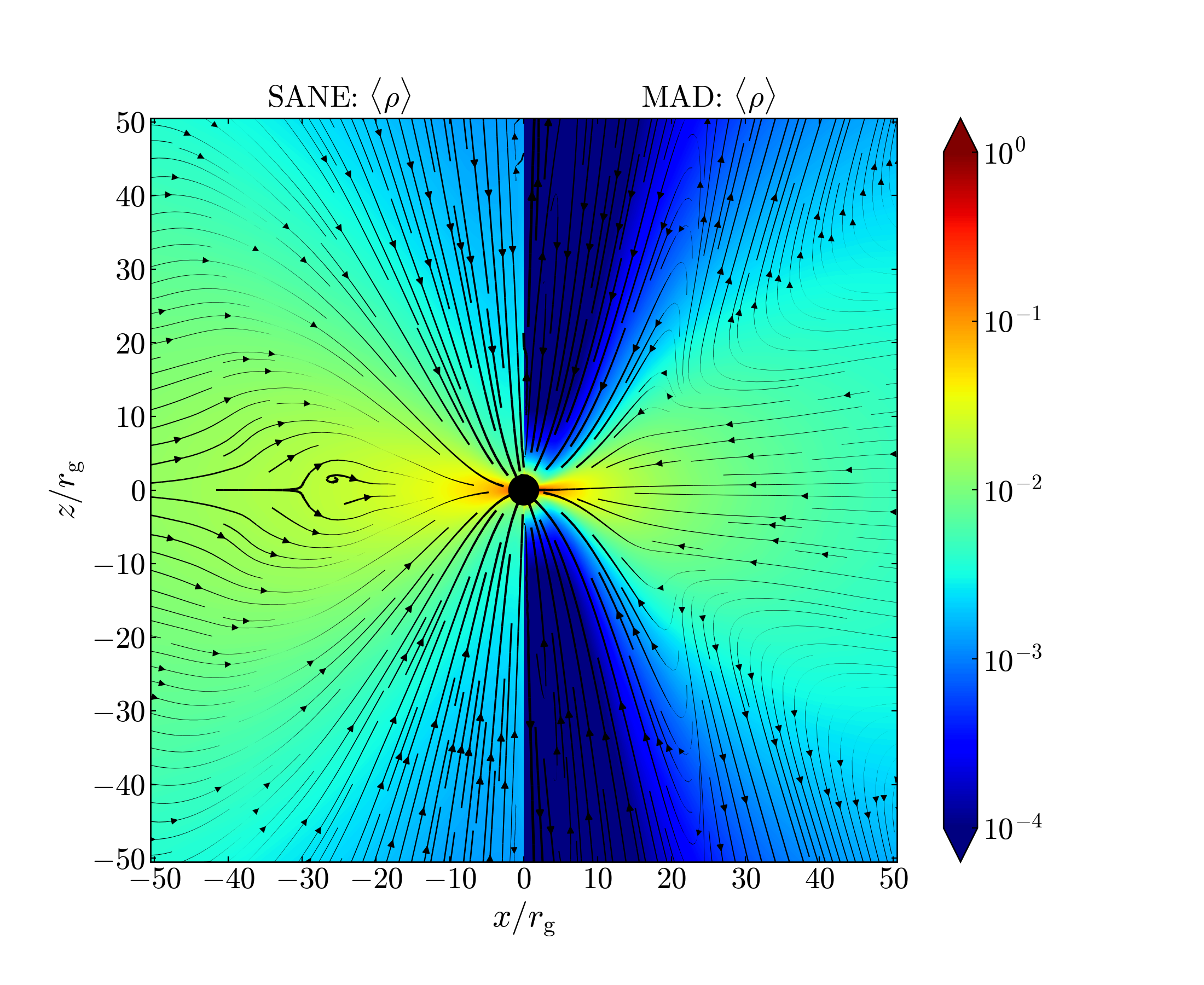}
	\caption{We show the gas density and velocity streamlines for the SANE (left side) and MAD (right side) accretion flows, time and azimuthally-averaged over the final $50000\tg$ for each simulation. We also symmetrize the data in the $\theta$-direction across the disk midplane. We indicate the magnitude of the velocities using the linewidths of the streamlines: light (bold) lines indicate small (high) velocities. Outgoing streamlines in the MAD model indicate a prominent wind component.}
    \label{fig:density_time_avg}
\end{figure}

\section{Simulation setup} \label{sec:code}

\noindent We use the GPU-accelerated GRMHD code \hamr{} \citep{liska_hamr2020_arxiv} to evolve the gas density, velocity, temperature and magnetic field over time. \hamr{} assumes a fixed Kerr spacetime, which is reasonable given the relatively short time evolution of our simulations as compared to the long timescales of black hole spin and mass evolution. We use logarithmic Kerr-Schild coordinates, i.e., $\{X^0,X^1,X^2,X^3\}\equiv \{t, \log\, r, \theta, \varphi\}$ and adopt geometrical units, $GM_{\rm BH}=c=1$, which normalize the gravitational radius $\rg=GM_{\rm BH}/c^2=1$ and the light-crossing time $t_{\rm g}=GM_{\rm BH}/c^3=\tg=1$. Our 3D simulation grid extends from $r=1.71r_{\rm g}$ to $10^3r_{\rm g}$. We have an effective resolution of $N_{r}\times N_{\theta}\times N_{\varphi}\equiv 580\times 288\times 512$. We use 1 level of external static mesh refinement (SMR) and 4 levels of internal SMR to reduce the $\varphi-$resolution to 32 cells for $0^{\circ}<\theta<3.25^{\circ}$, 64 cells for $3.25^{\circ}<\theta<7.5^{\circ}$, 128 cells for $7.5^{\circ}<\theta<15^{\circ}$, 256 for $15^{\circ}<\theta<30^{\circ}$, and the full $N_{\varphi}=512$ for $30^{\circ}<\theta<90^{\circ}$ \citep[see][for more details about SMR in \hamr{}]{liska_hamr2020_arxiv}. We use outflowing radial boundary conditions (BCs), transmissive polar BCs and periodic $\varphi$-BCs \citep[][]{liska_tilt_2018}. 

We initialize our simulation with a Schwarzschild black hole surrounded by a standard ``FM'' \citep{fis76} torus, taking the torus inner edge at $r_{\rm in}=20\rg$ and the pressure maximum at $r_{\rm max}=41\rg$. The maximum gas density is normalized to 1. For the gas thermodynamics, we assume an ideal gas equation of state with the gas pressure $p_{\rm gas}=(\gamma_{\rm ad}-1)u_{\rm gas}$ where $u_{\rm gas}$, is the internal energy and the adiabatic index $\gamma_{\rm ad}=13/9$. 

We performed two simulations, one that leads to a weakly magnetized accretion flow  \citep[denoted as ``SANE'';][]{Narayan:2012,Porth:19}, and the other a magnetically arrested disk \citep[or ``MAD'';][]{igu03,narayan03,tch11}. We initialize a single poloidal magnetic loop by applying a purely toroidal magnetic vector potential, $A_{\varphi} \propto \max(q,0)$. The expression for $q$ for the SANE and MAD simulations are: 

\begin{equation}
{\rm SANE: }\, q = \frac{\rho}{\rho_{\rm max}} - 0.2 \,,
\end{equation}
\noindent and, 
\begin{equation}
{\rm MAD: }\, q = \frac{\rho}{\rho_{\rm max}}\left(\frac{r}{r_{\rm in}}\right)^3 \sin^3 \theta \exp \left(-\frac{r}{400}\right) - 0.2 \,,
\end{equation}

\noindent respectively, where $\rho$ is the rest-mass gas density. The magnetic field strength in the initial setup is normalized by setting $\max(p_{\rm gas})/\max(p_{\rm mag})=100$, where $p_{\rm mag}=b^2/2$ is the magnetic pressure and $b$ is the co-moving frame magnetic field strength. To avoid numerical errors in the vacuous polar funnel, we inject density and internal energy in the drift frame \citep[][]{ressler_2017} whenever the magnetization $b^2/\rho c^2$ exceeds $20$.

\begin{figure}
    \includegraphics[width=\columnwidth]{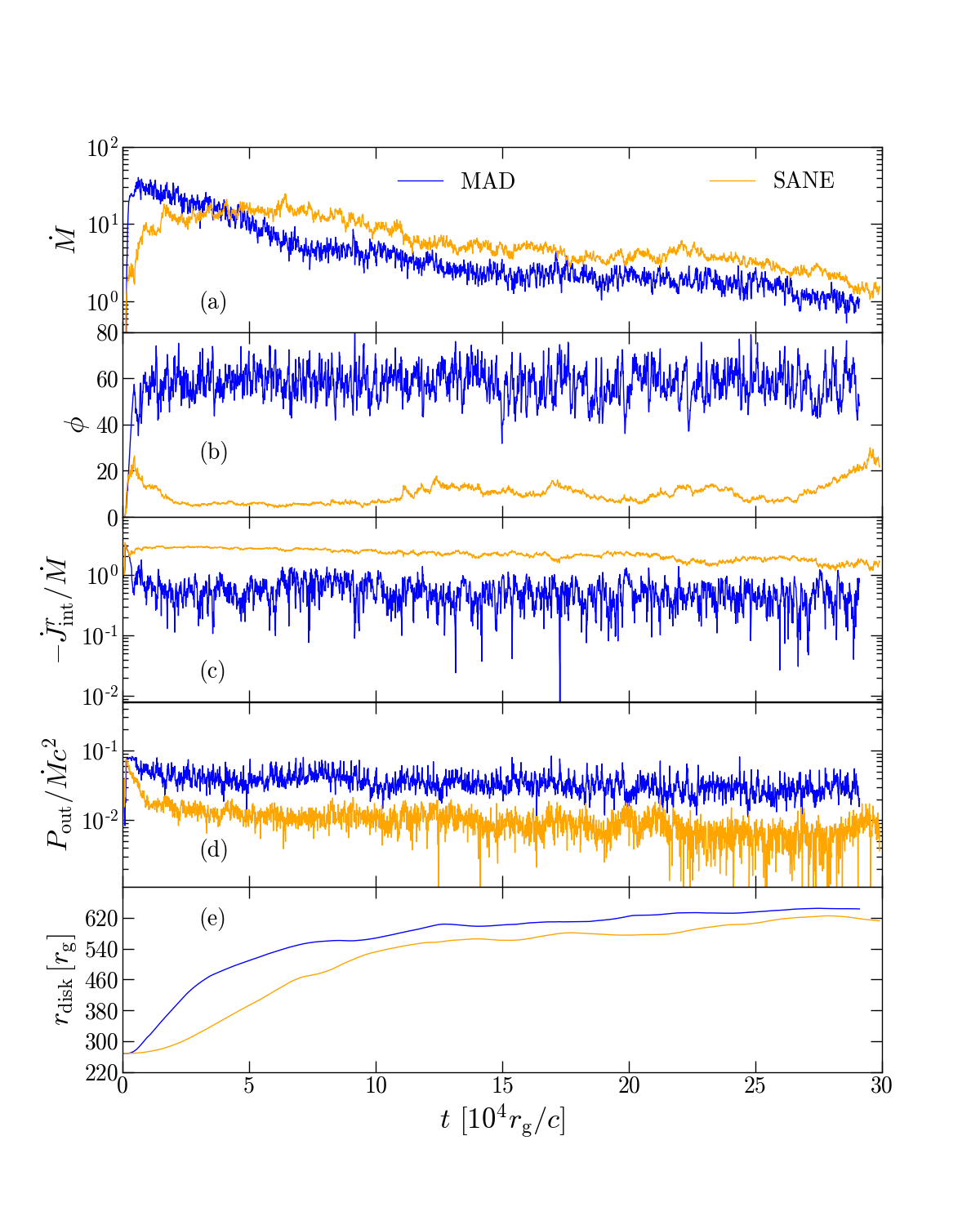}
	\caption{Weakly magnetized (SANE, orange) and magnetically arrested disks (MAD, blue) evolve quite differently over time. We show the time evolution of fluxes in the SANE and the MAD simulations: the accretion rate $\dot{M}$, dimensionless magnetic flux $\phi$, specific radial flux of the angular momentum $\dot{J}^r_{\rm int}/\Mdot$, outflow power efficiency $P_{\rm out}/\Mdot c^2$ and the disk barycentric radius $r_{\rm disk}$. We calculate $\Mdot$, $\Jtot$ and $P_{\rm out}$ at $5\rg$ in order to avoid any spurious effects from density floors. The magnetic flux is estimated at the event horizon.}
    \label{fig:time}
\end{figure}

\begin{figure}
    \includegraphics[width=\columnwidth]{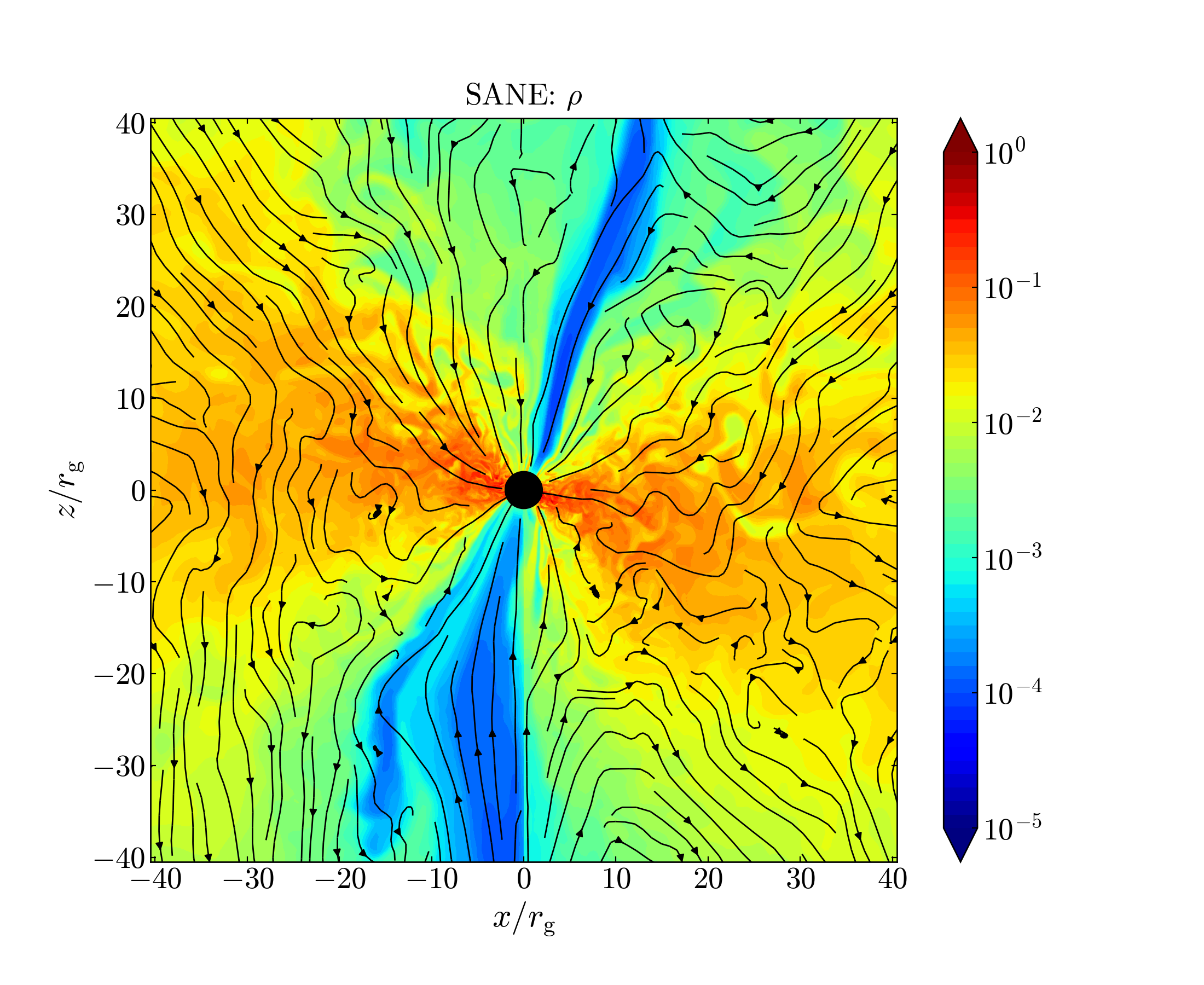}
	\caption{We see gas infall in the polar regions of the SANE simulation at late times. We show the gas density $\rho$ at $2\times10^5\tg$ with black lines indicating velocity streamlines. The disk also undergoes a change in orientation at times as large-scale eddies accrete at random times during the simulation.}
    \label{fig:SANE_winds}
\end{figure}

\section{Results} \label{sec:results}

\noindent Both simulations were evolved to a time $t\sim3\times10^5\tg$. The long runtime enabled us to reach inflow-outflow equilibrium out to $r_{\rm eq} \sim 100-150 \rg$ (discussed in Sec.~\ref{sec:radial}). Figure~\ref{fig:SANE_slice} shows the vertical and midplane cross-sections of the gas density $\rho$ and the ratio of gas and magnetic pressures, namely the plasma-$\beta \equiv p_{\rm gas}/p_{\rm mag}$, of the SANE accretion flow at $t=1.3\times 10^5 \tg$. There are no relativistic outflows, as is expected to be the case for a non-spinning black hole, and thus we see that gas is plunging towards the BH in the evacuated polar region. We also see turbulent disk winds propagate outwards (green regions in the vertical plot of density in Fig.~\ref{fig:SANE_slice}). The midplane cross-section shows that the inspiralling gas exhibits a laminar structure punctuated by small-scale eddies throughout the disk body, best seen in the plasma-$\beta$ plots. 

Figure~\ref{fig:MAD_slice} shows the vertical and midplane cross-sections of the MAD (magnetically arrested disk) model at $t=2.9\times 10^5\tg$. This model shows a wider polar region with prominent disk winds while the inner disk (within $\sim 10\rg$) is vertically much thinner compared to the SANE disk. The midplane cross-section shows that the infalling gas is disrupted by regions of density depression that also exhibit low plasma-$\beta$, indicating strong magnetic fields. These features are due to magnetic flux eruptions that occur when a magnetic flux bundle containing strong vertical fields escapes from the BH's magnetosphere and propagates radially outward into the disk \citep[e.g.,][]{tch11, Ripperda2022}. These features are in sharp contrast to the SANE disk, which exhibits many more turbulent eddies in the bulk flow as is expected from a MRI-dominated accreting gas \citep[e.g.,][]{Narayan:2012, Porth:19}. Accretion in MADs occurs via magnetic Rayleigh-Taylor/interchange instabilities \citep[e.g.,][]{igumenshchev08} as disk gas moves inwards by displacing the strong vertical fields. One other interesting feature to note is that the orientation of the disk can change over time: we began with a disk whose angular momentum vector was parallel to the z-axis, while Fig.~\ref{fig:MAD_slice} shows that the disk angular momentum vector at $t=2.9\times10^5\tg$ is slightly misaligned with respect to the vertical axis. We expect such misalignments in the accreting flow to be random in nature and subject to the formation and accretion of large scale eddies in the bulk of the disk.

Figure~\ref{fig:density_time_avg} shows a comparison of the time, azimuthally averaged, $\theta-$symmetrized density and velocity between the SANE and MAD models. The time-averaging is done over $t\approx240000-290000\tg$. The boldness of the streamlines indicate the velocity magnitude. The MAD model shows a thinner inflow region compared to the SANE model. The velocity streamlines in the MAD model vanish at the disk-wind boundary as inflowing streamlines turn outwards into a prominent wind component. This feature is absent in the SANE model within at least $50\rg$. The velocities are largest in the polar region of both models as gas free-falls towards the BH.

\subsection{Time evolution} \label{sec:time}

\noindent Next we study how the disks in the two models change over our long simulation run time. Figure~\ref{fig:time} shows the long term evolution of the mass accretion rate $\dot{M}$, shell-integrated total angular momentum flux in the radial direction $\dot{J}^r_{\rm int}$ , total outflow power $P_{\rm out}$ (all calculated at $r=5\rg$) and the disk barycentric radius $r_{\rm disk}$. We also calculate the dimensionless magnetic flux $\phi$ (in Gaussian units) at the event horizon radius. We take $\dot{M}$ and $\dot{E}$ to be positive for mass and energy inflow towards the BH, while $\dot{J}^r_{\rm int}$ is positive for angular momentum outflow. These quantities are defined as: 
\begin{equation}
    \dot{M} = -\iint \rho u^r \, \! \sqrt{-g}\, d\theta \, d\varphi \,, \\ [10pt]
\end{equation}
\begin{equation}
    \dot{J}^r_{\rm int} = \iint T^r_{\varphi} \, \! \sqrt{-g}\, d\theta \, d\varphi \,, \\ [10pt]
\end{equation}
\begin{equation}
   P_{\rm out} = \dot{M}c^2-\dot{E}\,, \\ [10pt]
\end{equation}
\begin{equation}
   {\rm where}\,  \dot{E} = \iint T^r_t \, \! \sqrt{-g}\, d\theta \, d\varphi \,, \\ [10pt]
\end{equation}
\begin{equation}
    \phi = \frac{\sqrt{4\pi}}{2\sqrt{\dot{M}}}\iint |B^{r}| \, \sqrt{-g}\, d\theta \, d\varphi\,, \\ [10pt]
\end{equation}
\begin{equation}
    r_{\rm disk} = \dfrac{\iiint r \, \rho \,  \sqrt{-g}\, dr\, d\theta \, d\varphi}{\iiint \rho \, \sqrt{-g}\, dr\, d\theta \, d\varphi}\,.
\end{equation}

\noindent Here $g\equiv|g_{\mu\nu}|$, $u^r$, $B^r$, $T^{r}_{t}$ and $T^{r}_{\varphi}$ are the metric determinant, the radial components of the 4-velocity and the 3-magnetic field, and the radial fluxes of the energy and angular momentum, respectively:
\begin{eqnarray}
    T^r_t=(\rho + \gamma_{\rm ad} u_{\rm g}+b^2)u^ru_t - b^rb_t, \\
    T^r_{\varphi}=(\rho + \gamma_{\rm ad} u_{\rm g}+b^2)u^ru_{\varphi} - b^rb_{\varphi}.
\end{eqnarray}

In both the MAD and SANE simulations, $\dot{M}$ at $5\rg$ (Fig.~\ref{fig:time}, panel a) becomes quasi-steady for $t>10^5 \tg$. The total mass within the simulation grid slowly decreases with time as gas flows into the BH and outflows from the disk remove gas beyond the outer grid boundary. For the MAD simulation, the horizon dimensionless magnetic flux saturates at around $60$ and is punctuated by sharp dips due to magnetic flux eruptions \citep[see e.g.,][]{tch11}. The value of $\phi-$saturation of 60 is larger than the nominal value of $50$ for non-spinning BHs \citep[e.g.,][]{Narayan2022} possibly because of the higher spatial resolution employed in the present study. For the SANE case, the horizon magnetic flux hovers between 5 and 15, which is an indication of the presence of weakly magnetized accretion. At late times, the SANE simulation exhibits larger $\phi$ values ($\sim 25$) perhaps due to the polar infall of magnetized gas, though never coming close to reaching the saturation value of 60. The specific radial flux of the total angular momentum, $\dot{J}^r_{\rm int}/\Mdot$ stays roughly constant over time in the SANE disk. This quantity varies rapidly in the MAD case dropping by an order of magnitude at times. The average value of $\dot{J}^r_{\rm int}/\Mdot$ for the MAD model ($\sim 0.45$) is lower than for the SANE model ($\sim 1.69$), suggesting highly sub-Keplerian rotation. Interestingly, the specific angular momentum flux for the MAD case exhibits several dips similar to those in the magnetic flux, though not necessarily at the same time. However it suggests a connection between the two features. 

We also see similar dips in the MAD outflow power $P_{\rm out}$, which is, on average, $\sim 5\%$ of the inflowing accretion power $\Mdot c^2$. For the SANE model, the outflow power is $\lesssim 1\%$ of the accretion power. Since there is no jet in either model, all of the outflow power comes in the form of slow-moving gas-rich winds. Indeed, due to their low power, the winds are unable to prevent the disk midplane from shifting out of the equatorial plane, which has important consequences for the evolution of the SANE model. At early times, $t \lesssim 10^5 \tg$, there is an evacuated polar region roughly perpendicular to the SANE disk midplane (see Fig.~\ref{fig:SANE_slice}). As eddies in the SANE disk get tossed about by large-scale turbulence (at $r \gtrsim 100 \rg$), the polar region gets filled in over time. This results in a quasi-spherical accretion structure at late times (see Fig.~\ref{fig:SANE_winds}). In the MAD case, polar infall is prevented as the relatively stronger wind maintains a coherent structure over time. These results indicate the need to run simulations for a long time as these large-scale eddies are only formed and accreted over very long timescales. Indeed, as we see from the barycentric radius $r_{\rm disk}$ of the disk (Fig.~\ref{fig:time}), the viscous spreading of the disk continues to be significant until about $t\sim 2\times 10^5\tg$, indicating that the bulk of the disk only achieves quasi-steady state beyond this time. 

\begin{figure}
    \includegraphics[width=\columnwidth]{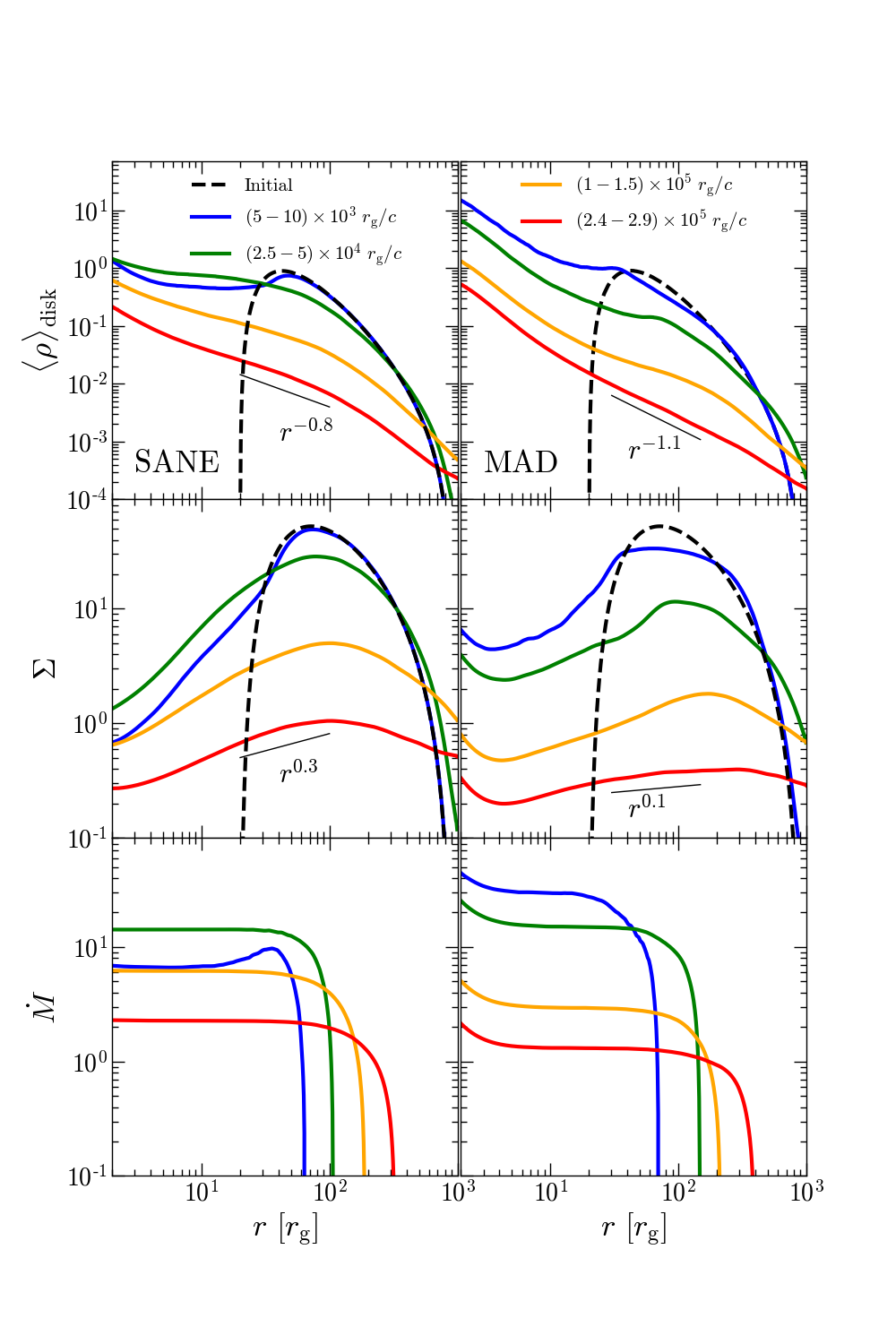}
	\caption{There is significant evolution of the radial disk structure over time in both models. We show the time-averaged disk gas density $\rho$, surface density $\Sigma$ and mass accretion rate $\Mdot$, calculated for four time chunks (identified by color). By the final time chunk, each model achieves inflow-outflow equilibrium out to at least $100-150\rg$, as seen from the $\Mdot$ profiles. Radial power-law fits to $\rho$ and $\Sigma$ are shown over a radius range where the disk scale aspect ratio is roughly constant in the two models (see Fig.~\ref{fig:disk_radial}).}
    \label{fig:disk_radial_time}
\end{figure}

\subsection{Radial disk structure} \label{sec:radial}

\begin{figure}
    \includegraphics[width=\columnwidth]{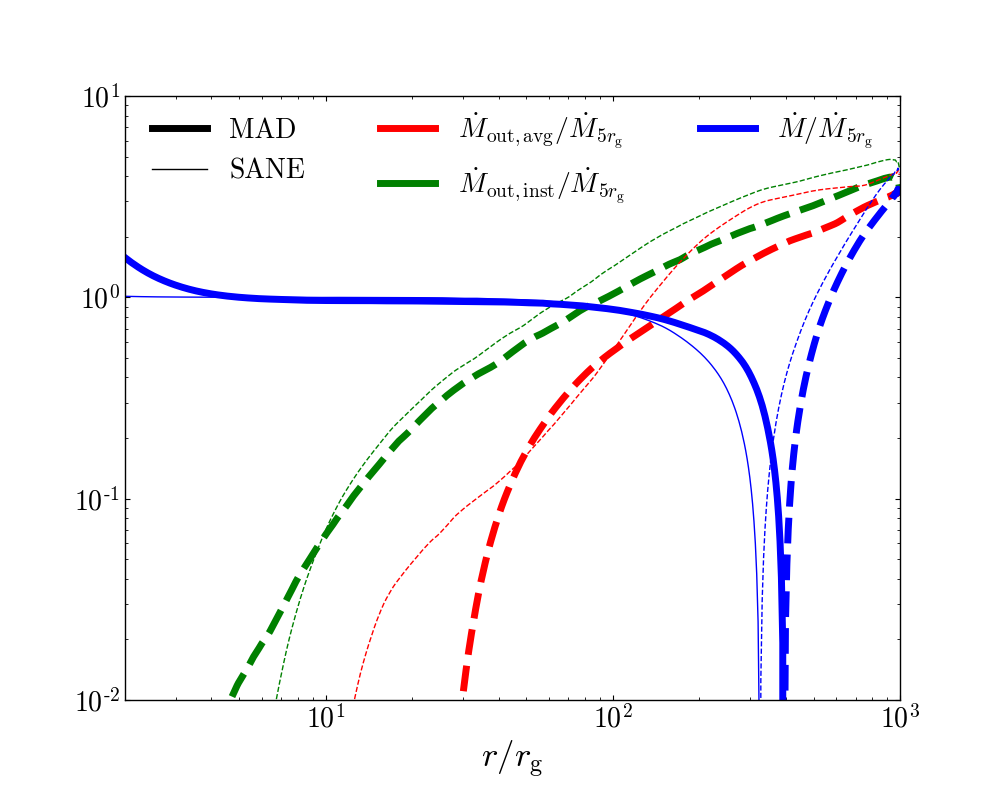}
	\caption{We compare the normalized net inflow mass-accretion rate $\Mdot$ and the mass outflow rate $\Mdot_{\rm out}$ for the MAD (thick lines) and SANE (thin lines) models, time-averaged over the final $5\times 10^4\tg$ of time. Inward (outward) fluxes are indicated by solid (dashed) lines. We perform two types of calculations for the mass outflow rate: $\dot{M}_{\rm out, avg}$ where we apply constraints on the $t,\varphi$-averaged outward radial velocity ($v_r>0$) and specific radial energy flux $(\mu_{\rm e}>0)$, and $\dot{M}_{\rm out, inst}$ where the constraints are applied on the instantaneous values of $v_r$ and $\mu_{\rm e}$ at each time. These two calculations provide lower and upper bounds on the true $\Mdot_{\rm out}$. Both models achieve an average outflow rate between $60-80\%$ of the net mass inflow rate at $r\approx 100\rg$, showing that winds do not efficiently remove gas from the disk.}
    \label{fig:Mdot_radial}
\end{figure}

As we saw in the previous section, the properties of the accretion disk change significantly over time. Here we study the radial profiles of a variety of disk properties, time-averaged over different chunks of the simulation time. Figure~\ref{fig:disk_radial_time} shows the change of the radial profiles of the disk-averaged gas density $\rho$, surface density $\Sigma=(1/2\pi)\iint \rho rd\theta d\varphi$ and the mass accretion rate over time. We calculate the disk-averaged quantities $\langle q \rangle$ using the following equation:  
\begin{equation}
    \langle q \rangle_{\rm disk} = \frac{\iint q \rho \sqrt{-g} d\theta d\varphi}{\iint \rho \sqrt{-g} d\theta d\varphi}.
    \label{eq:disk_avg}
\end{equation} 

We choose four time chunks: $5000-10000\tg$, $25000-50000\tg$, $100000-150000\tg$ and $240000-290000\tg$, over which we time-average the quantities. The disk gas density decreases over time in the two simulations, with the radial slopes steepening to $-0.8$ and $-1.1$ for SANE and MAD respectively. One notable feature is that the disk density peak, which is initially at $41\rg$, shifts to larger radii as the inner part of the disk accretes on to the BH and the outer disk spreads out. The slope transitions from shallow to steep as we cross this ``peak,'' with the transition becoming smoother over time. It is only in the case of the final time chunk in the MAD model that the slope becomes roughly constant over the entire disk. 

For $\Sigma$, the slope in the inner accretion flow gradually becomes shallower as time increases, with the final time chunk showing slopes of 0.3 and 0.1 for the SANE and MAD models, respectively. Such a decrease in the absolute value of $\Sigma$ was also noted in \citet{Narayan:2012} though the slopes were roughly constant over time in their models, possibly because the disk did not viscously spread as much due to the lower grid resolutions of those simulations. Indeed, \citet{liska_tilt_2018} and \citet{Porth:19} noted that disk spreading is vastly different when MRI is not well resolved at large radii, especially in the case of weakly magnetized disks. 

\begin{figure}
    \includegraphics[width=0.49\columnwidth]{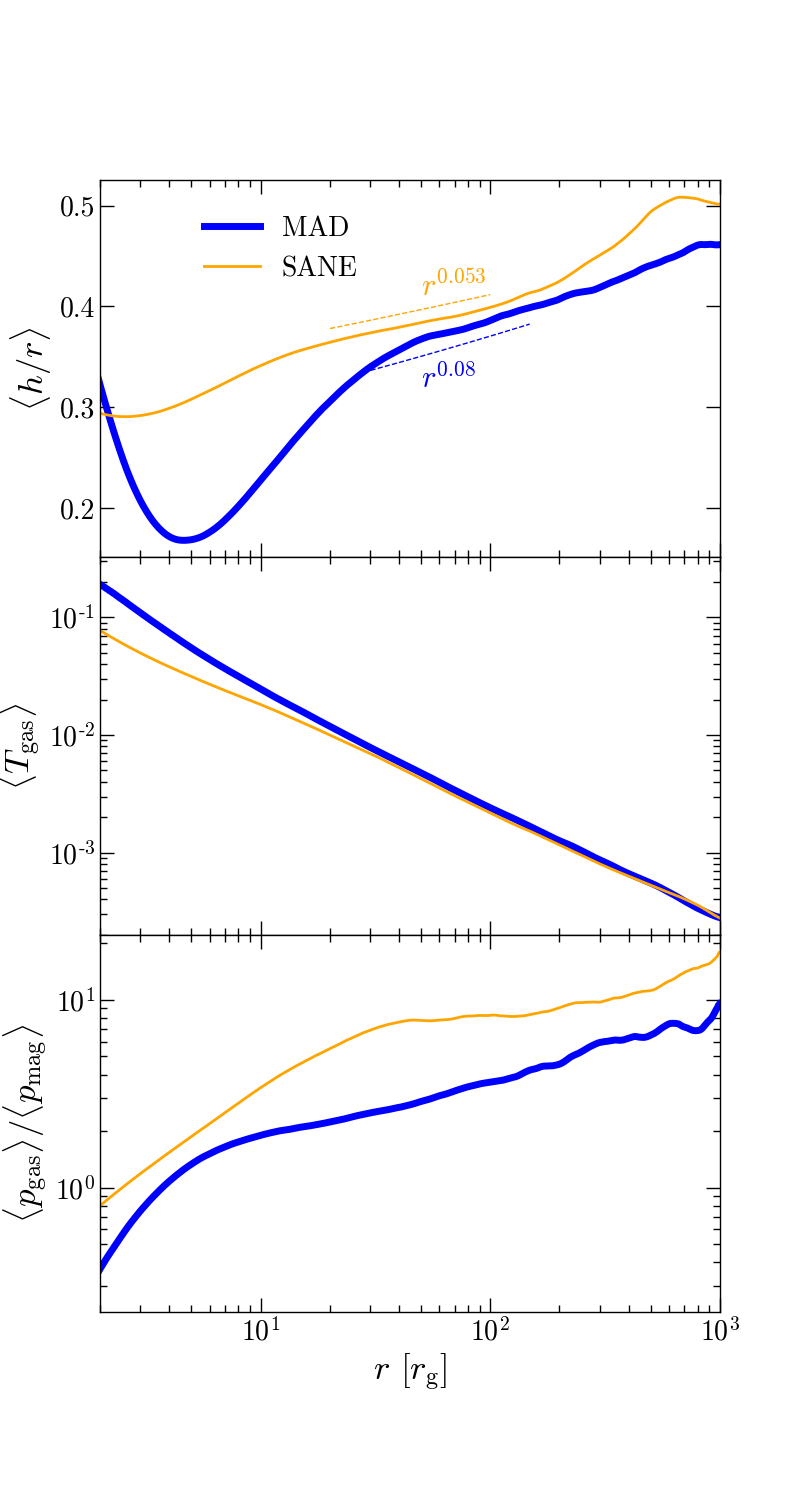}
    \includegraphics[width=0.49\columnwidth]{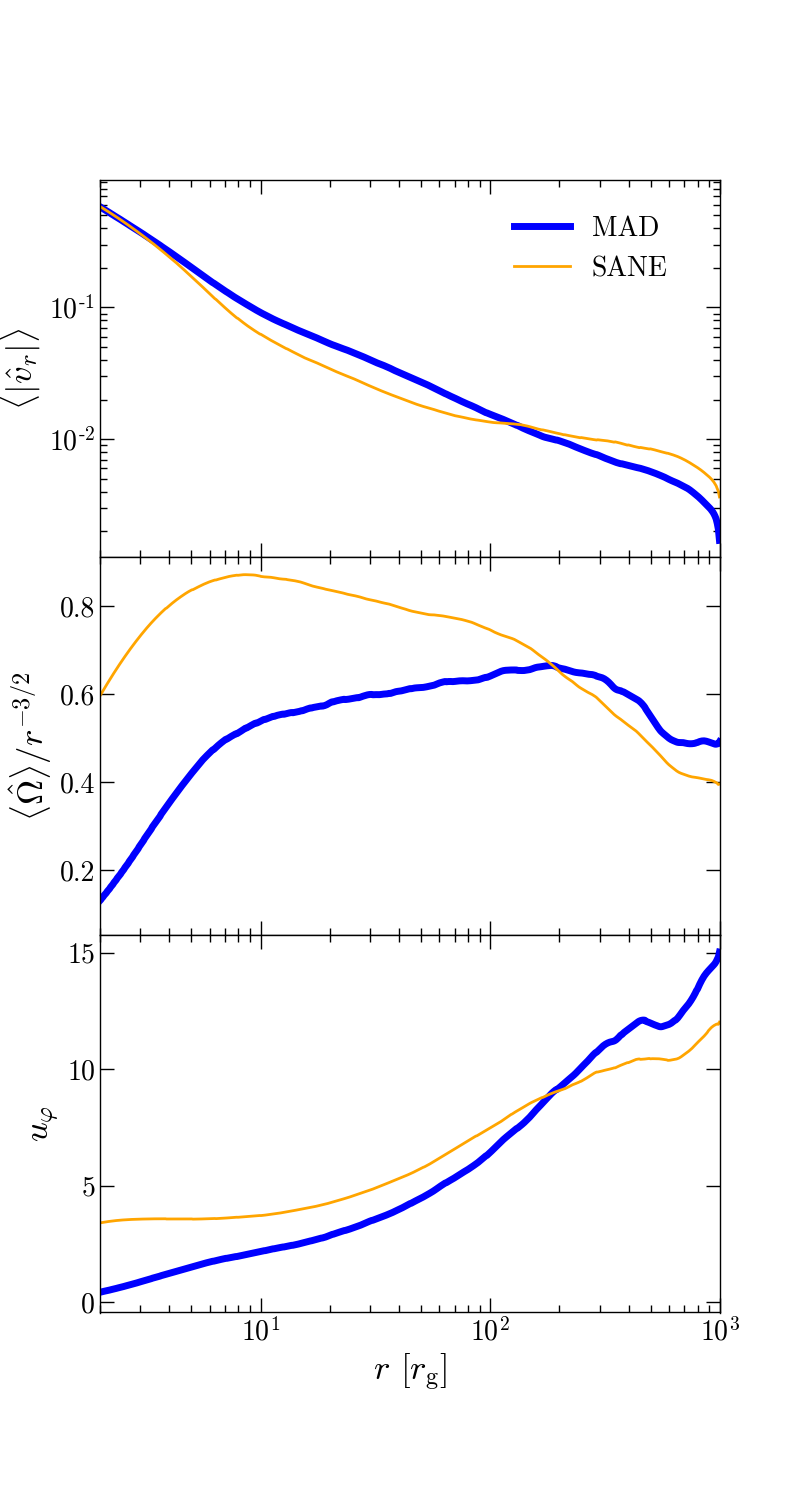}
	\caption{We show the radial structure of the disk scale aspect ratio ($h/r$), gas temperature $T_{\rm gas}$, plasma-$\beta \, (=p_{\rm gas}/p_{\rm mag})$, radial velocity $|v_r|$, angular velocity $\Omega$ and specific angular momentum $u_{\varphi}$. All quantities are disk-averaged. The MAD model has a thinner (i.e., smaller $h/r$), more magnetized (smaller $\beta$), hotter (larger $T_{\rm gas}$) and more sub-Keplerian (smaller $\Omega$, $u_\phi$) inner disk compared to the SANE model. The time-averages are performed over the final $5\times 10^4\tg$ of time for each simulation.}
    \label{fig:disk_radial}
\end{figure}

We expect that $\Sigma\sim \rho h$, where $h$ is the disk scale height. We fit the radial profiles of $\rho$ and $\Sigma$ between $r\sim 20-100\rg$ in the SANE model and $r\sim30-150\rg$ in the MAD model. We chose these radial ranges for the fit since the disk scale aspect ratio $h/r$ is roughly similar between the models in this region as we will see later in Fig.~\ref{fig:disk_radial}. Using our fits of $\rho$ and $h/r$, we expect $\Sigma$ slopes of $\sim0.25$ and $0.02$ for the SANE and MAD models. Comparing with the actual $\Sigma$ from Fig.~\ref{fig:disk_radial_time}, we see that the expected and fitted slopes match very well for the SANE model and they are also fairly similar for the MAD model. This suggests that the disks at the chosen radii have become radially self-similar. We discuss the implications of the radial slopes of $\rho$ for Sgr A$^*$ and M87$^*$ in Sec.~\ref{sec:density_slope}.

\begin{figure}
    \includegraphics[width=\columnwidth]{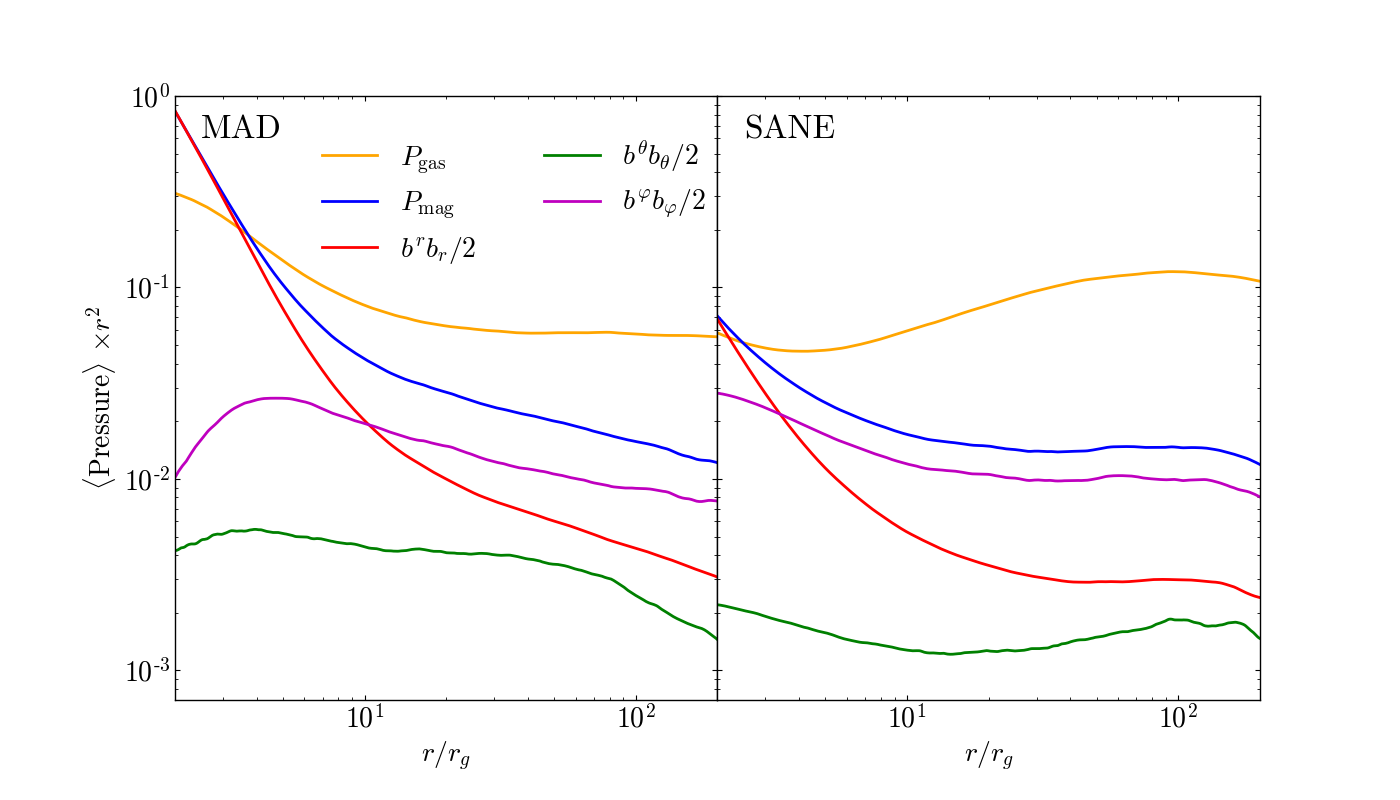}

	\caption{The MAD disk is, on average, magnetic pressure dominated within the inner few gravitational radii while the SANE disk is always thermal pressure supported. We show the radial profiles of the time- and disk-averaged thermal and magnetic pressure as well as the different components of the magnetic pressure. The radial component of the magnetic pressure of the MAD model exceeds the toroidal component, indicating that the inner disk contains strong poloidal fields. The time-averaging is done over the same period as in Fig.~\ref{fig:disk_radial}.}
    \label{fig:pressure}
\end{figure}

Figure~\ref{fig:disk_radial_time} also shows that the mass accretion rate $\Mdot$ changes sign at roughly the peak of the surface density. Inside this peak radius, $\Mdot$ is constant and we deem the accretion flow to be in inflow-outflow equilibrium. The radial range over which inflow-outflow  equilibrium is achieved increases substantially between the first and last time chunk. This is the motivation for our choice of evolving our models to very long timescales (as in the original work of \citealt{Narayan:2012}). 

Figure~\ref{fig:Mdot_radial} shows that the radial profiles of the time- and shell-averaged mass accretion rate for the two simulations behave similarly, with the simulations achieving inflow-outflow equilibrium out to $r\sim100-150\rg$ (total $\dot{M}$ is constant within this radius). The increase in $\Mdot$ inside $3\rg$ for the MAD model is probably due to accretion of artificially floored gas density. We also show two different calculations of the outward mass flux $\dot{M}_{\rm out}$, depending on how we determine mass loss via winds. 

First we have $\dot{M}_{\rm out, avg}$, where we only account for the mass flux in regions that exhibit an outward time- and $\varphi-$averaged radial velocity in addition to a positive $t,\varphi-$averaged specific radial energy flux $\mu_{\rm e}$, which we define as 
\begin{equation}
    \mu_{\rm e}=-\frac{T^r_t}{\rho u^r}-1.
\end{equation}
This definition of mass outflow flux is the same as that given in \citet{Narayan:2012} where the idea was to determine whether the gas element is able to escape to infinity when accounting for its averaged properties over a long time period. 

We also calculate the instantaneous mass outflow flux, denoted by $\dot{M}_{\rm out, inst}$, where for each instantaneous snapshot of the simulation, we count any gas that has outward-oriented radial velocity and sufficient energy to escape ($\mu_e>0$) to be part of the outflow. This method of calculating the mass outflow rate is similar to that used in \citet{Yuan:12_winds,Yuan:15}. Generally, $\dot{M}_{\rm out, inst}$ is larger than $\dot{M}_{\rm out, avg}$.

We see that magnitudes of both $\dot{M}_{\rm out, avg}$ and $\dot{M}_{\rm out, inst}$ are small compared to the net accretion rate inside the inflow-equilibrium radius. This is especially true close to the BH ($r\lesssim10\rg$) where even the instantaneous mass loss efficiency is less than $10\%$ of the net $\dot{M}$. The average outflow rates $\dot{M}_{\rm out, avg}$ reach $60-80\%$ of $\dot{M}$ at around $100\rg$, i.e., winds are not yet dominant, in agreement with the results in \citet{Narayan:2012}. The instantaneous $\dot{M}_{\rm out, inst}$ is larger, perhaps up to $\sim 2\Mdot$.  Overall, it appears that disk winds around Schwarzschild black holes are weak and turbulent in nature, with gas moving out and then rejoining the inflow at larger radii. These results are fairly consistent with the behavior of winds seen in \citet{Yuan:12_winds} and \citet{Yuan:15}, where the authors separated out the turbulent mass outflow flux and the real outflow by tracking velocity trajectories, and found mass loss efficiencies close to $200\%$ at $r \sim 80\rg$ \footnote{We note that the instantaneous mass outflow rate shown in \citet{Yuan:15} is somewhat larger than what we find, possibly due to a different initial disk setup.}. 

Figure~\ref{fig:disk_radial} shows the radial profiles of the disk scale aspect ratio $h/r=\langle|\theta-\pi/2|\rangle_{\rm disk}$, gas temperature $T_{\rm gas}=p_{\rm gas}/\rho$, plasma-$\beta$, radial velocity $v_r$, angular velocity $\Omega$ and the angular momentum $u_{\varphi}$, all disk-averaged (as in eq.~\eqref{eq:disk_avg}) and time-averaged over the final $5\times 10^4\tg$ for each simulation. Further, for all the quantities except $h/r$, we average only over one scale height either side of the disk midplane. The inner disk region of a MAD flow is, on average, very different from that of a SANE model. The magnetic field strength in the BH magnetosphere for a MAD is so dominant (with disk plasma-$\beta \sim 2-3$ within a few tens of $\rg$) that the polar field lines push down vertically on the inflowing gas, thus increasing the disk gas density while lowering the disk scale aspect ratio. Our values of plasma-$\beta$ are larger than that found in some other works because of our choice of averaging the gas and magnetic pressure separately. \citet{Ressler:2021} averaged $\beta^{-1}$ over the disk, thereby preferentially weighting highly magnetized regions of the disk. Comparing the two approaches, our calculation of $\beta$ provides an upper limit while the \citet{Ressler:2021} method provides a lower limit. 

The squeezing of the inflow close to the BH leads to multiple magnetic reconnection events that result in hotter gas \citep[e.g.,][]{Ripperda2022}, as seen from the gas temperature $T_{\rm gas}$ in the inner $10~\rg$. Indeed, within this region, both gas and magnetic pressures are larger for the MAD disk as compared to the SANE model, as seen in Fig.~\ref{fig:pressure}. Even though the bulk of the disk for both cases has a toroidally-dominated magnetic field pressure, the radial component becomes larger close to the BH, especially in the MAD model where $b^rb_r/b^{\varphi}b_{\varphi}\sim2-10$ within $\sim10\rg$. The large radial magnetic pressure vertically supports the MAD disk near the BH, thereby causing $h/r$ to increase close to the BH. 

\begin{figure*}
	\includegraphics[width=\textwidth]{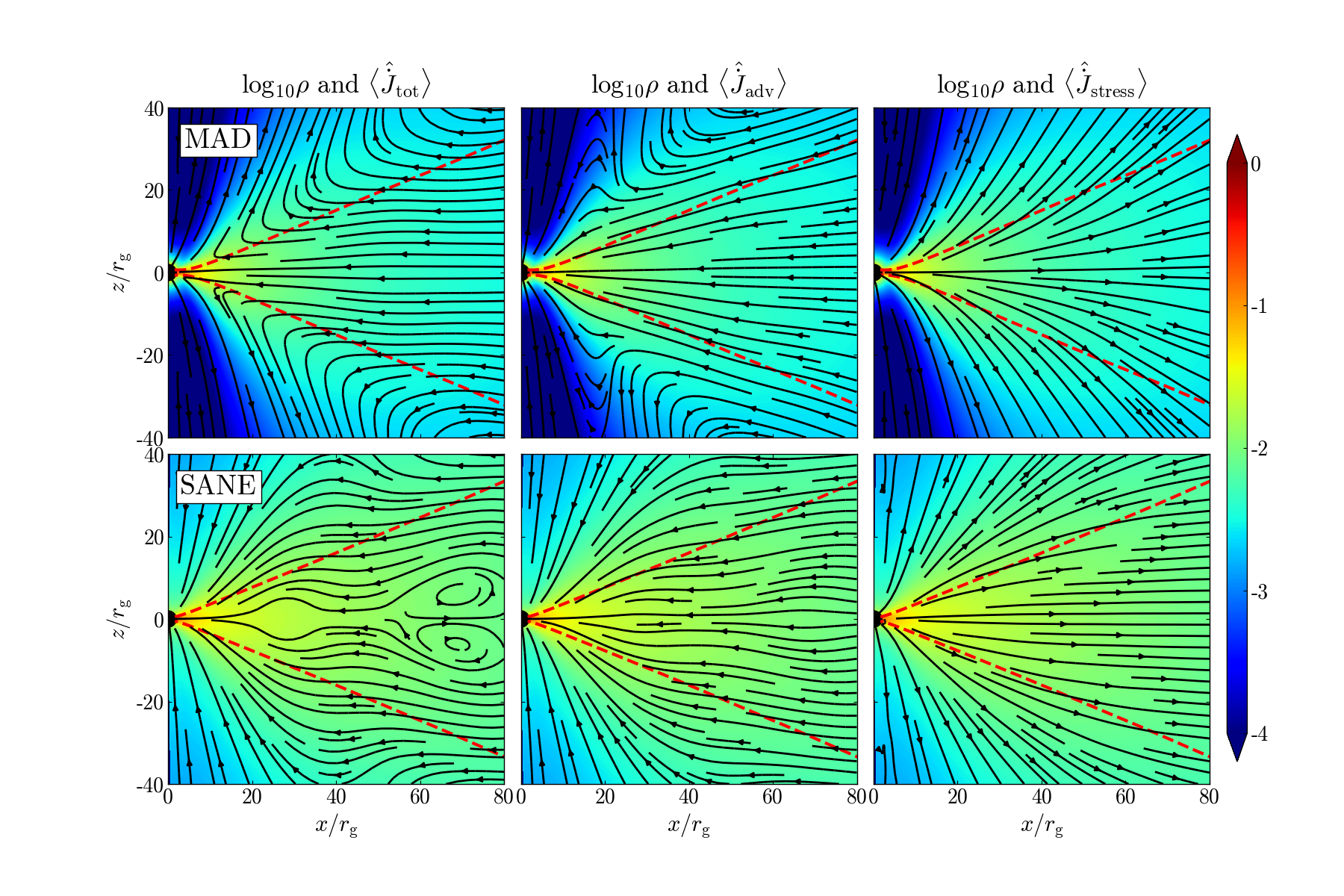}
    \caption{We show streamlines of the total angular momentum flux $\Jtot$, the (mostly inward) advective flux $\Jadv$ and the (outward) stress-induced flux $\Jout=\Jtot-\Jadv$, for the MAD (top) and SANE (bottom) simulations, with gas density in color. All quantities were $t,\varphi$-averaged and $\theta$-symmetrized in order to reduce the effects of turbulent eddies. Larger-scale eddies still persist in the SANE model as seen from the $\Jtot$ streamlines at $r>60\rg$. The disk scale height is indicated with red dashed lines (see Fig.~\ref{fig:disk_radial}), which demarcate the disk region, outside of which disk winds can transport angular momentum outwards if strong enough. On average, the outward stress-induced angular momentum transport in the MAD model is more vertically-oriented while in the SANE model, the outward flux is generally more equatorial in nature (compare the panels in the right column, and also see Fig.~\ref{fig:radial_separation}).}
    \label{fig:Jdot_2D}
\end{figure*}

Returning to Fig.~\ref{fig:disk_radial}, we plot the physical components (denoted by the hat symbol) of the velocities, i.e. $\hat{v}_i=v^i\sqrt{g_{ii}}$. The radial velocity is roughly similar between the SANE and MAD models. We note that $v_r$ for the SANE model in \citet{Narayan:2012} exhibits a steeper slope than our SANE model at larger radii while $v_r$ matches well for the MAD models. The discrepancy in the SANE $v_r$ is largely due to the increase in disk magnetization of the SANE model over time ($\phi$ increases from 10 to $\sim25$ between $1-3\times10^5\tg$). Standard SANE disks in the literature \citep[e.g., ][]{Narayan:2012, Porth:19, EHT_M87_2019_PaperV} exhibit $\phi$ values closer to 5-10. It is also possible that this discrepancy in the velocity partially arises due to stronger turbulence at large radii in our models, similar to the $\Sigma$ profile in Fig.~\ref{fig:disk_radial_time}.However, stronger turbulence does not seem to result in a larger outward mass flux at the outer boundary of the disk in our SANE model as indicated by the similarity of $\dot{M}_{\rm out, avg}$ from our models compared to those in \citet{Narayan:2012}. In the case of the angular velocity $\Omega$, our choice of a FM torus initially provides us with a super-Keplerian angular velocity profile inside the disk pressure maximum radius (at $41\rg$) and a sub-Keplerian profile beyond it. As time passes, for the MAD model, the strong vertical fields pinch the incoming accretion flow via reconnection, resulting in the ejection of magnetic flux bundles from near the event horizon \citep[see][]{Ripperda2022}. These flux-tubes move outward and interact with the accreting gas, reducing the flow angular velocity $\Omega$ to highly sub-Keplerian values ($\sim 0.2-0.6~\Omega_{\rm K}$) as well as decreasing the average specific angular momentum $u_{\varphi}$ as compared to SANE disks.

\section{Angular momentum transport} \label{sec:jtot}

\begin{figure}
    \includegraphics[width=\columnwidth]{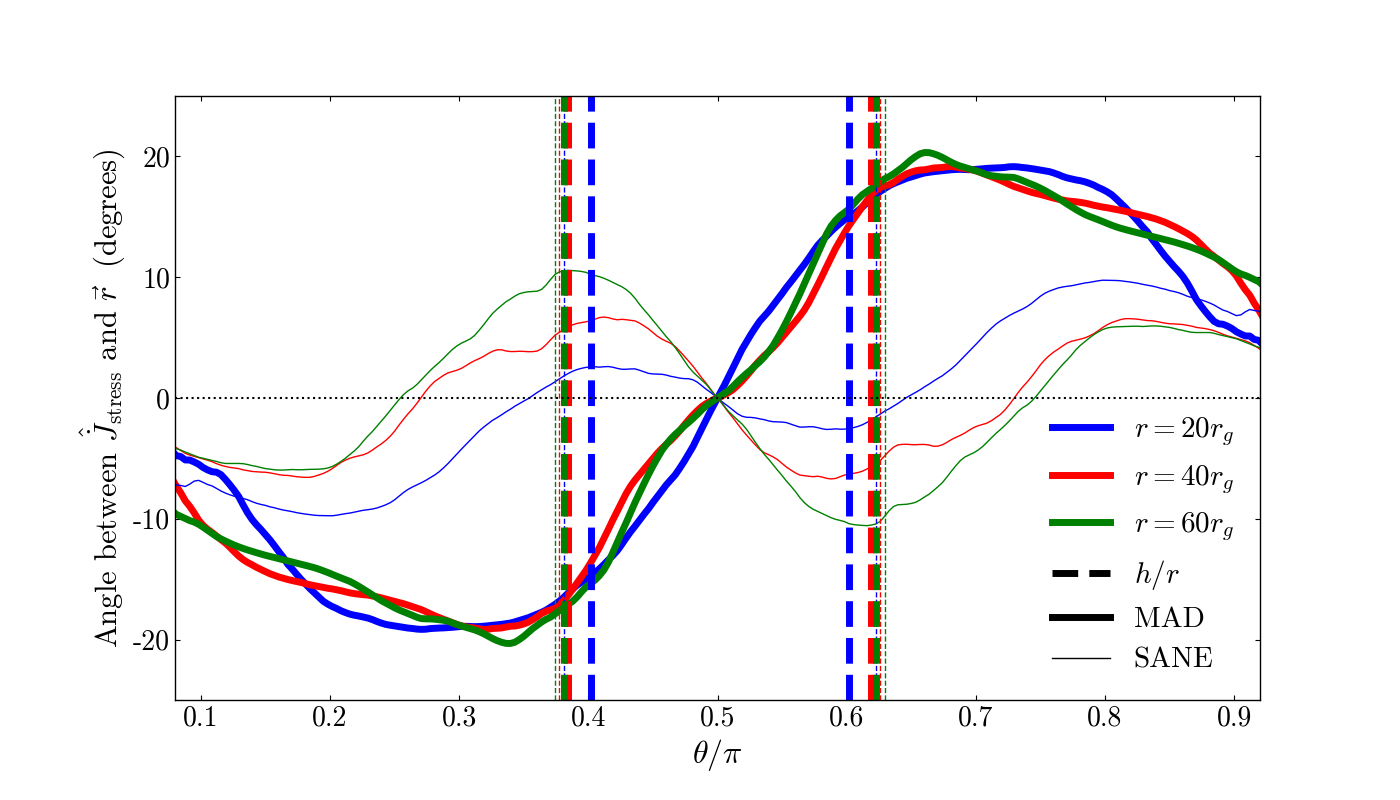}

	\caption{We show the angle between the poloidal vector $\Jout^i$ and the radial vector $\Vec{r}$ at three different radii: 20, 40 and $60\rg$. For the MAD model, the angular separation is negative, with $\Jout$ reaching a maximum offset of $\sim20^{\circ}$ counterclockwise (clockwise) from $\Vec{r}$ in the upper (lower) hemisphere of the wind region. The sense of the deviation indicates vertical outward transport of angular momentum. We see the opposite behavior in the SANE model, which is consistent with horizontal transport of angular momentum. The vertical dashed lines indicate the disk scale aspect ratio for both models at each radius.}
    \label{fig:radial_separation}
\end{figure}

\noindent In this section, we calculate the angular momentum flux $\Jtot$ for the MAD and SANE models, separating out the inward advective, and the outward stress-induced or ``viscous,'' parts of the flux. Following the approach in \citet{Narayan:2012}, we axisymmetrize and time-average the total angular momentum flux,
\begin{equation}
    \dot{J}^i_{\rm tot}(r,\theta) = \left\langle T^i_{\varphi}\right\rangle_{\varphi, t},
\end{equation}
where $i \equiv r,\, \theta$, the symbol $\langle \cdots \rangle_{\varphi, t}$ indicates an average over azimuthal angle and time, and 
\begin{equation}
T^i_{\varphi}=(\rho + \gamma_{\rm ad} u_{\rm g}+b^2)u^i u_{\varphi} - b^i b_{\varphi}.
\end{equation}
For the advective component of the angular momentum flux $\Jadv$, we adopt the definition given by \citet{penna10}, where the authors took the product of the mean velocities, $\langle u^r \rangle $ and $\langle u_{\varphi} \rangle $ as part of the ``in-going'' angular momentum flux, placing the correlated fluctuations in $\langle u^r u_{\varphi} \rangle $ as a contribution to the transport due to Reynolds stresses. Thus, we have

\begin{equation}
    \dot{J}^i_{\rm adv}(r,\theta)=\left\langle\left(\rho+u_{\rm g}+\frac{b^2}{2}\right) u^i\right\rangle_{\varphi, t} \left\langle u_{\varphi} \right\rangle_{\varphi, t}.
    \label{eq:Jadv}
\end{equation}

\noindent Note that we have taken $b^2/2$ to be part of the advective component as this is the contribution of the magnetic field to the energy density of the gas, and plays a role similar to $u_g$. Both these contributions to the energy density, along with the rest mass density $\rho$, are advected with the gas flow \citep[also see][]{penna10}.

We perform the time-averaging for $\Jtot$ and $\Jadv$ over the final $5\times10^4\tg$ of each simulation. Further, to get rid of the effects of small-scale turbulent eddies in the disk, we symmetrize $\Jtot$ and $\Jadv$ in the $\theta-$direction, accounting for the direction of the flux in each hemisphere, i.e., radial components of the fluxes are symmetrized across the midplane while polar components are anti-symmetrized. Once we have the $t,\varphi-$averaged, $\theta-$symmetrized structure of $\Jtot$ and $\Jadv$, we calculate the outward angular momentum flux due to fluid stresses as simply 
\begin{equation}
    \Jout=\Jtot - \Jadv.
\end{equation}
Figure~\ref{fig:Jdot_2D} shows the streamlines of the different angular momentum fluxes for the SANE and MAD models. From this point, we only discuss the physical components (i.e., ``hatted'') of angular momentum flux, i.e., $\hat{\dot{J}}^i=\dot{J}^i\sqrt{g_{ii}}$, so we drop the hat for brevity. 

First we focus on the $\Jtot$ streamline morphology. For the SANE model, at $r \gtrsim60\rg$, we see the effects of large-scale turbulent disk eddies that still linger even after averaging over $5\times10^4\tg$. Within $r=60\rg$, the streamlines are roughly radially flowing inwards and seem to become more equatorial as we transition from the polar region to the disk. The SANE disk wind is too weakly powered to show any significant amount of outward $\Jtot$, and thus $\Jtot$ seems to be always inflowing at least within $r\sim 80\rg$. 

In the MAD model, the $\Jtot$ streamlines are much more uniform as compared to the SANE model as we have inflow-outflow equilibrium out to at least $100\rg$ (see Fig.~\ref{fig:Mdot_radial}). Within the disk, i.e., inside one scale height either side of the midplane as shown by the red dashed lines in Fig.~\ref{fig:Jdot_2D}, we see similar equatorial flux transport as the SANE model. There is a change from inflowing to outflowing streamlines as we move from the disk to the wind. Thus, despite the absence of a jet, the disk wind in MADs is strong enough to enable outflow of angular momentum flux. 

Moving on to the angular momentum flux due to advection $\Jadv$, we generally see inward advective flux in both models, except for a small region in the MAD disk wind. In the SANE model, $\Jadv$ is entirely inward directed as the advective flux transitions from equatorial to radial inflow as we move from the disk midplane to the poles ($\theta=0$ and $\pi$) since we essentially have free-falling gas in the polar region. The $\Jadv$ streamlines match the pattern of the velocity streamlines shown in Fig.~\ref{fig:density_time_avg}.

\begin{figure}
    \includegraphics[width=\columnwidth]{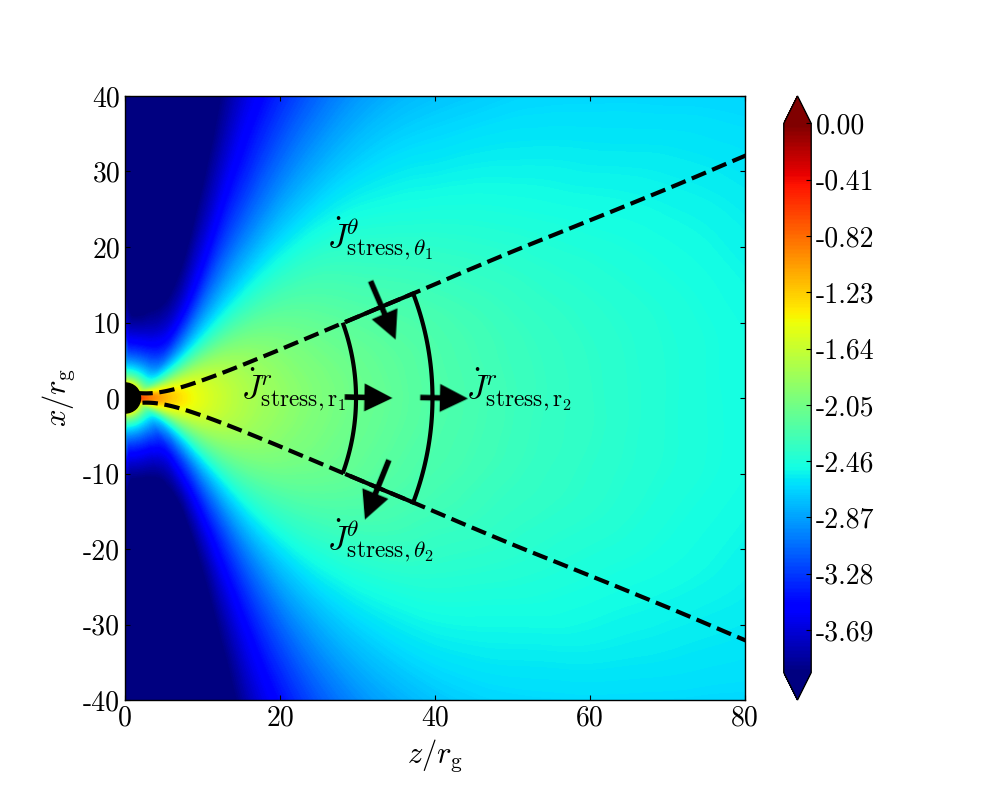}
    \includegraphics[width=\columnwidth]{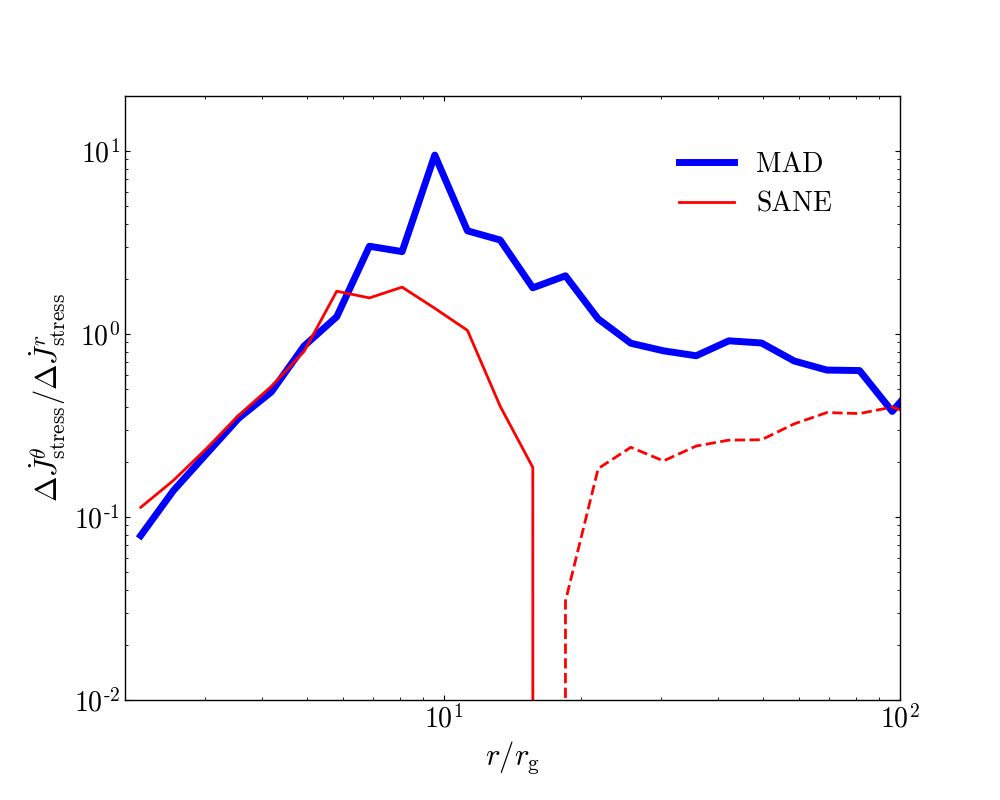}
	\caption{The MAD simulation exhibits an outward vertical flux of angular momentum caused by stresses, while SANE disks are dominated by equatorial flux. Top: a schematic diagram showing the calculation of the net $\Jout$ for an annulus within the disk. The black dashed lines indicate $h/r$. Bottom: We show the ratio of the net outward angular momentum flux due to fluid stresses in the polar and radial directions. Solid (dashed) lines correspond to net positive (negative) ratio, and indicate the direction of the net polar flux. In the MAD model, the outward polar flux is as strong as the outward radial flux, consistent with the result in Figure.~\ref{fig:radial_separation}. In contrast, the SANE model has a weak polar influx of angular momentum at $r\gtrsim20\rg$.}
    \label{fig:net_Jout}
\end{figure}

The right column of Fig.~\ref{fig:Jdot_2D} shows the stress-induced or ``viscous'' angular momentum flux $\Jout$. There is a clear indication of a difference in the orientations of the streamlines between the SANE and MAD models. This difference is most obvious when we consider streamlines that cross the dashed red lines indicating the disk scale height. With increasing radius, in the MAD disk the streamlines move from below the scale height to above, whereas the opposite occurs in the SANE disk. The change in orientation of the streamlines is better seen in Fig.~\ref{fig:radial_separation} where we quantify the deviation from a purely radial structure by calculating the angle between the vector $\Jout$ and the radial vector $\Vec{r}$. Positive (negative) values of the angle indicate a clockwise (counterclockwise) shift from the radial vector. We see that the SANE $\Jout$ vector maintains a deviation $\lesssim 10^{\circ}$ for $r\leq60\rg$. Within $20^{\circ}$ of the midplane, $\Jout$ is essentially equatorial since it is clockwise shifted from the radial vector in the upper hemisphere and counterclockwise shifted in the lower hemisphere. The MAD disk, on the other hand, exhibits a $\Jout - \Vec{r}$ angular separation pattern with the opposite sign. Here $\Jout$ is more vertically oriented relative to the radial vector. The magnitude of the angular deviation is also larger than in the SANE model.

To gauge the relative importance of the outward polar transport compared to the radial transport, we calculate the net rate of outflow of angular momentum from the annulus of the disk shown in Fig.~\ref{fig:net_Jout}. For this annulus, the radial outflow of angular momentum is described by
\begin{eqnarray}
    \Delta \dot{J}_{\rm stress,\, (r_1,r_2)}^r = &\int^{2\pi}_{0}\int^{\theta_2}_{\theta_1}\Jout^r \sqrt{-g}d\theta \, d\varphi|_{r_2} \nonumber \\
    &-\int^{2\pi}_{0}\int^{\theta_2}_{\theta_1}\Jout^r \sqrt{-g}d\theta \, d\varphi|_{r_1}\,,
\end{eqnarray}

\noindent where the two integrals are computed at the inner and outer edges, i.e., $r=r_1$ and $r_2$, of the disk annulus. The fluxes crossing the top and bottom edges of the annulus are,
\begin{equation}
    \dot{J}_{\rm stress,\, \theta_2}^{\theta} = -\dot{J}_{\rm stress,\, \theta_1}^{\theta} = \int^{2\pi}_{0}\int^{r_2}_{r_1}\Jout^{\theta} \sqrt{-g}dr \, d\varphi,
\end{equation}
\noindent where the integration is done at $\theta=\theta_{(1,2)}=\pi/2 \pm h/r$ corresponding to one disk scale height on either side of the midplane. The values of $\dot{J}_{\rm stress,\, \theta_{1,2}}^{\theta}$ are equal but opposite in sign since we have anti-symmetrized the polar components of the fluxes across the midplane. Hence, the net polar outward flux through the annulus is 

\begin{equation}
    \Delta \Jout^{\theta}=2\times \dot{J}_{\rm stress,\, \theta_2}^{\theta}.    
\end{equation}

Figure~\ref{fig:net_Jout} shows the ratio of the polar and the radial components of the angular momentum outflow due to stresses. First, we note that the ratio of fluxes is negative for the SANE model, indicating that the polar flux is directed towards the midplane (as seen also in Fig.~\ref{fig:radial_separation}). The predominantly equatorial outflow of angular momentum in the SANE model aligns well with the notion that the MRI is the primary mechanism of angular momentum transport and, therefore, is restricted to the disk region. On the other hand, the MAD disk exhibits significant vertical outward flux with $\Delta \Jout^{\theta}/\Delta \Jout^r\sim 1$, suggesting that winds play a very important role in angular momentum transport.

\subsection{Decomposing $\Jout$ into Maxwell and Reynolds components} \label{sec:jmax}

\begin{figure}
	\includegraphics[width=\columnwidth]{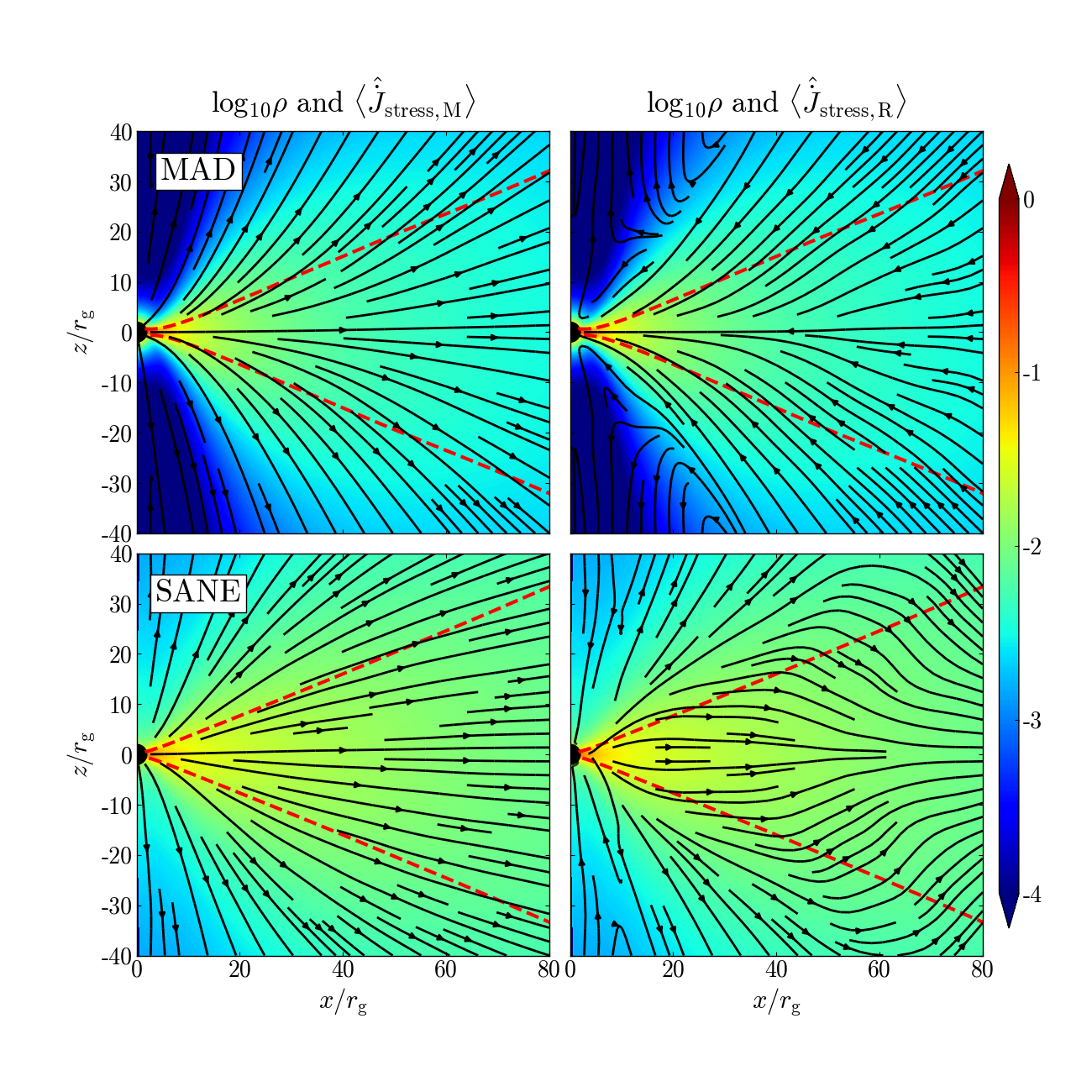}
    \caption{Same as Fig.~\ref{fig:Jdot_2D}, but for the Maxwell and Reynolds components of the stress-induced angular momentum flux, $\JoutM$ and $\JoutR$. The Maxwell component dominates and thus, the streamlines look similar to that of the total $\Jout$. An important feature shown here is that $\JoutR$ points inward for the MAD model while it points outward for the SANE model.}
    \label{fig:Jout_MR2D}
\end{figure}

\begin{figure}
    \includegraphics[width=0.45\textwidth]{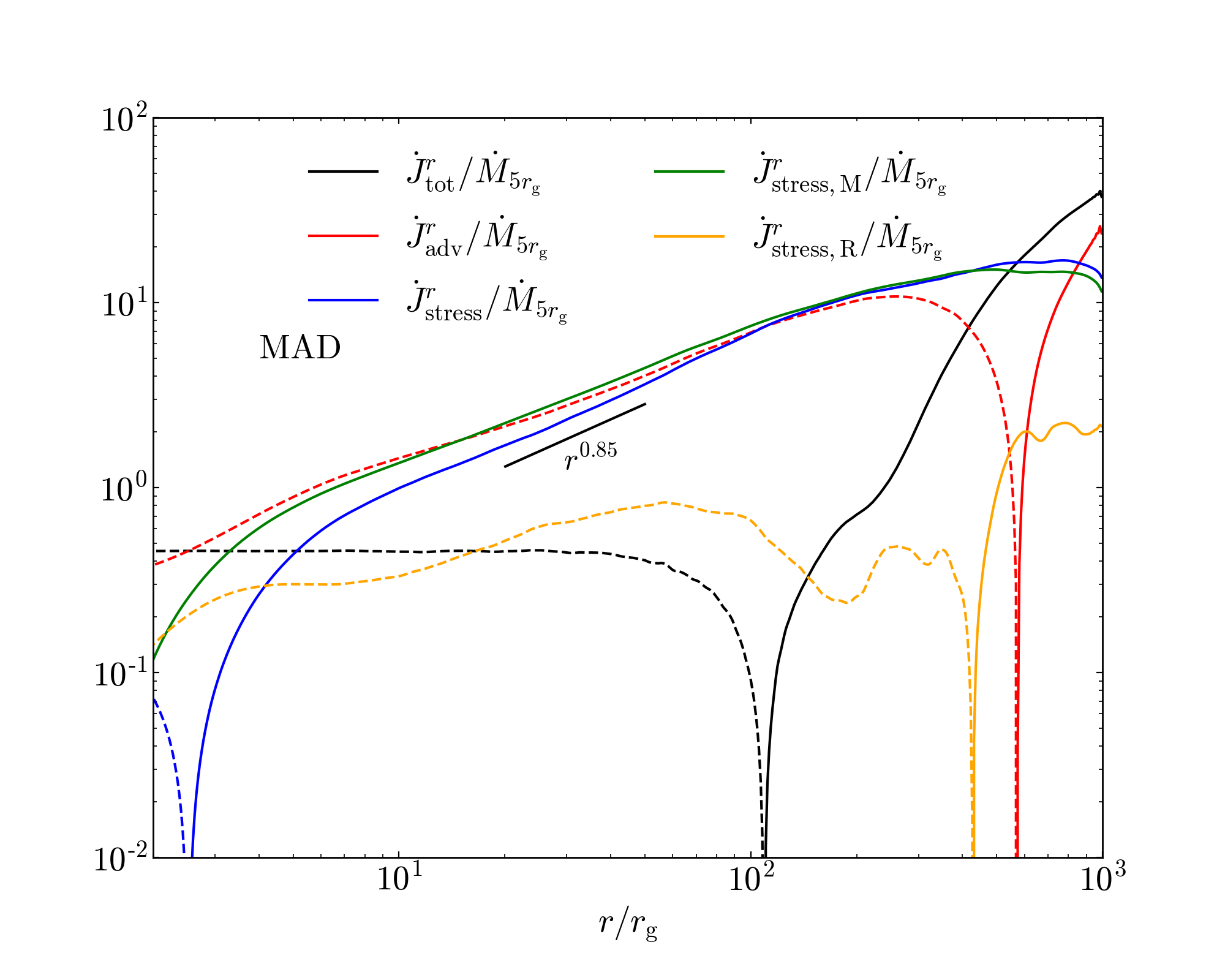}
    \includegraphics[width=0.45\textwidth]{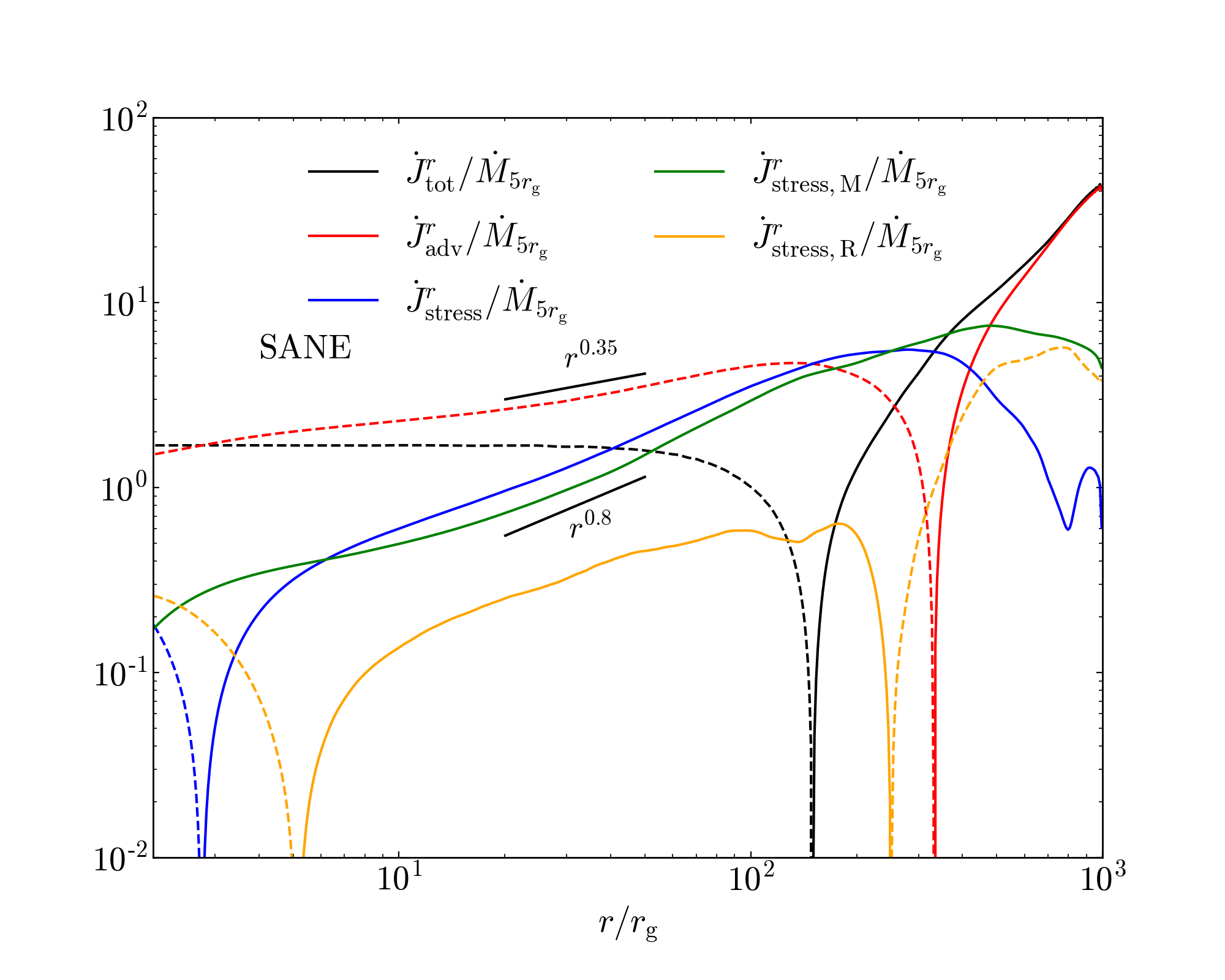}
	\caption{We show the the shell-integrated components of the radial flux of angular momentum (from Fig.~\ref{fig:Jdot_2D} and \ref{fig:Jout_MR2D}) for the MAD (top) and SANE (bottom simulations), time-averaged over the final $50,000\tg$. Solid (dashed) lines indicate outward (inward) transport. Maxwell stresses dominate the outward angular momentum flux for both models, but is so strong for the MAD model that it matches the inward advective flux, showcasing the importance of magnetic fields in regulating angular momentum transport in MADs. }
    \label{fig:Jdot_radial}
\end{figure}

\begin{figure}
    \includegraphics[width=\columnwidth]{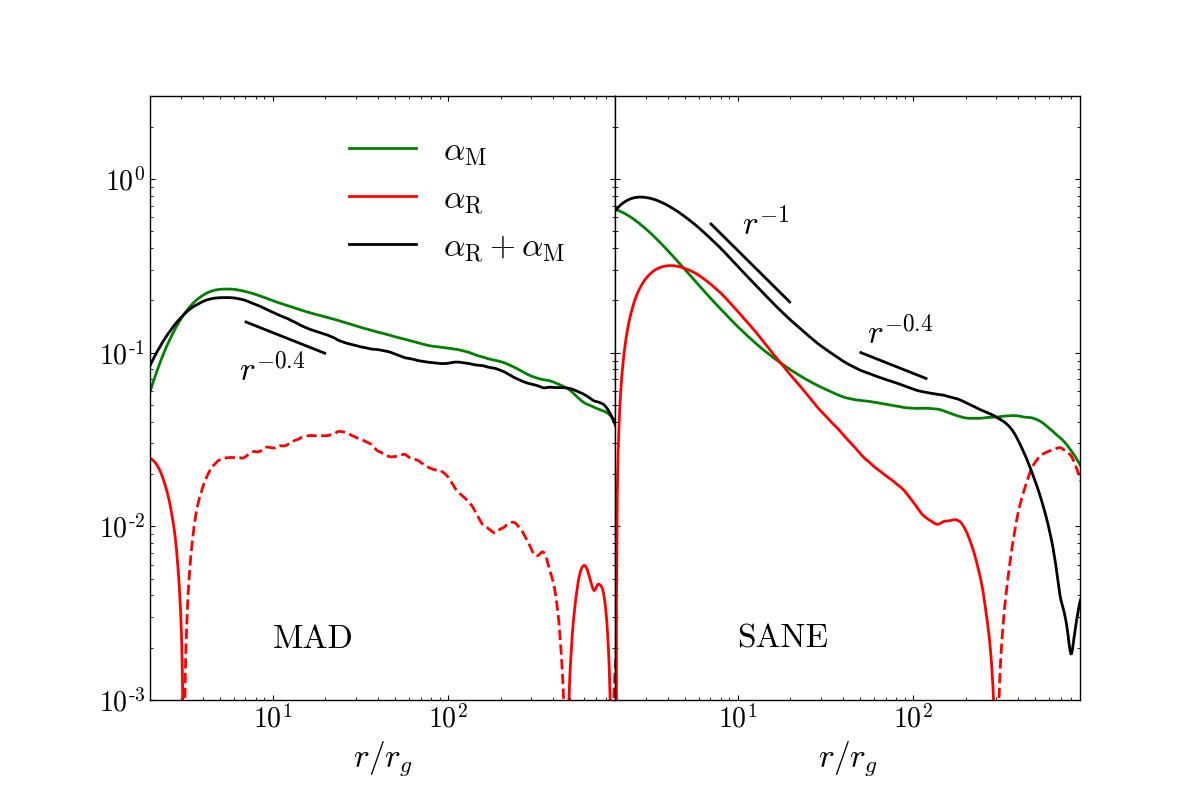}
	\caption{We show the time-averaged $\alpha$ viscosity parameters for both models: the Maxwell stress $\alphaM$, the Reynolds stress $\alphaR$ and the net value of $\alphaR+\alphaM$. Solid (dashed) lines indicate outward (inward) stresses. The MAD model exhibits an inward Reynolds stress, which results in inward transport of angular momentum as was seen in Fig.~\ref{fig:Jdot_radial}.}
    \label{fig:alpha_radial}
\end{figure}

\noindent Here we take a closer look at the stress-induced angular momentum flux $\Jout$, separating out the flux contributions due to the Maxwell and Reynolds stresses. The expression for the Maxwell component of the angular momentum flux is given by
\begin{equation}
    \dot{J}^i_{\rm stress, M}(r,\theta)= \left\langle\frac{b^2}{2} u^i u_{\varphi} - b^i b_{\varphi}\right\rangle_{\varphi, t}.
\end{equation} 

\noindent Here we include both the mean and the turbulent fluxes associated with the magnetic fields. For the contribution due to Reynolds stresses, we account for the correlated fluctuations associated with the gas:
\begin{equation}
    \JoutR^i(r,\theta)= \left\langle\left(\rho+u_{\rm g}+\frac{b^2}{2}\right) u^i u_{\varphi} \right\rangle_{\varphi, t} - \Jadv^i,
\end{equation}
\noindent where $\Jadv$ is given by eq.~\eqref{eq:Jadv}.

Figure~\ref{fig:Jout_MR2D} shows the structure of $\JoutM$ and $\JoutR$ for the SANE and MAD models. We see that the Maxwell component of $\Jout$ for each model looks very similar to the total $\Jout$, indicating that the Maxwell component dominates over the Reynolds component in both models and sets the direction of $\Jout$. The Reynolds component for the SANE model is driven by small-scale fluctuations whereas $\JoutR$ in the MAD model closely resembles its Maxwell counterpart. More importantly for the MAD model, the direction of the $\JoutR$ streamlines is opposite to that of $\JoutM$, suggesting that the Reynolds stresses are responsible for inward angular momentum transport. 

Figure~\ref{fig:Jdot_radial} shows the time-averaged and shell-integrated radial component of the angular momentum fluxes in the two simulations. We achieved a constant $\dot{J}^r_{\rm tot}$ up to $\sim 80-100~\rg$ in both simulations, and therefore, we can conservatively say that the disks have reached quasi-steady-state within $80\rg$. Comparing the MAD and the SANE profiles for the accretion rate-normalized angular momentum fluxes, we see that the total radial flux of the angular momentum is larger in the SANE case (also seen in Fig.~\ref{fig:time}). This suggests that non-spinning black holes surrounded by a weakly magnetized disk accrete angular momentum much quicker (also see Sec.~\ref{sec:spinup}). 

Apart from the magnitude of $\Jtot$, there are two major differences in the flux profiles between the SANE and MAD models. One is the relative strength of $\Jout$. We see that $\Jout\sim\Jadv$ for the MAD model, while $\Jout$ is significantly smaller in the SANE model. The absolute values of $\Jout$ are similar between both models. This suggests that the magnetic stresses in the MAD disk is as efficient in removing angular momentum as MRI in the SANE disk. The other prominent difference between the models is the aforementioned change in sign of $\JoutR$. The Maxwell and Reynolds components have the same sign in the SANE model, but opposite signs in the MAD model. To verify the nature of the stresses in our models, we calculate the $\alpha$ viscosity coefficients due to the Maxwell ($\alpha_{\rm M}$) and Reynolds ($\alpha_{\rm R}$) stresses:

\begin{eqnarray}
\alpha_{\rm M}   &=& -\hat{b}_r\hat{b}_{\varphi}/(p_{\rm gas}+p_{\rm mag}), \\
\alpha_{\rm R}   &=& (\rho+\gamma_{\rm ad}u_{\rm gas}+b^2)\delta \hat{u}_r \delta \hat{u}_{\varphi}/(p_{\rm gas}+p_{\rm mag}).
\end{eqnarray}

\noindent Here, $\delta \hat{u}_i= \hat{u}_i - \langle \hat{u}_i \rangle_{\rm disk}$ are the turbulent components of the gas velocity. 

Figure~\ref{fig:alpha_radial} shows that $|\alphaR| << \alphaM$ near the BH for MAD, while these quantities are similar in magnitude for SANE. The same behavior is seen in Fig.~5 of \citet{liska_tor_2019}, where the authors show that a poloidal flux-deficient disk can develop large-scale poloidal loops via the so-called $\alpha-\Omega$ dynamo, and eventually transition to the MAD regime. We further note that in the MRI-dominated regime, the Maxwell and Reynolds components of the stresses have the same sign \citep[][]{Pessah:2006}, resulting in positive outward angular momentum transport contributions from both components. This is what we see in the SANE case. Note that we have absorbed the negative sign within $\alphaM$, so that the net viscosity is $\alphaR+\alphaM$, which is different to the notation used in \citet[][]{Pessah:2006}. In the MAD model, the time-averaged $\alphaR$ has the opposite sign to that of $\alphaM$, and therefore leads to inward angular momentum transport \citep[also see, e.g.,][]{narquataertigum02,igu03}. We also find that the net viscosity $\alphaR+\alphaM$ under-predicts the expected radial velocity $v_r=(3/2)\alpha c_{\rm s} (h/r)$ within the inner $\sim 100\rg$ when compared to the radial velocity values shown in Fig.~\ref{fig:disk_radial}. This discrepancy is especially prominent for the MAD model due to gas plunging inwards close to the BH. Here $c_{\rm s}= \sqrt{\gamma_{\rm ad}p_{\rm gas}/(\rho+u_{\rm gas}+p_{\rm gas})}$ is the sound speed.

 The inward $\JoutR$ and negative values of $\alphaR$ in the MAD model from Fig.~\ref{fig:Jdot_radial} and \ref{fig:alpha_radial} suggest convection-like behavior in the MAD model \citep[see e.g., ][]{Begelman2022}. While we do not explicitly address convective instabilities in MADs in this work, it is possible that convection manifests in the form of sheared flux-tubes that propagate out as buoyant magnetic bubbles, often seen in disrupted jets \citep[see Sec.~\ref{sec:MADtime} and, e.g.,][]{Ressler:2021, Kaaz:2022}. From previous studies, it has been shown that MADs are at least marginally convectively-unstable \citep[][]{Narayan:2012,Begelman2022} but the relatively low values of $\JoutR$ that we find in our study indicate that the flux due to turbulent convection is subdominant. In this case, convection due to fluid turbulence should be relatively unimportant in MADs, except when heating occurs due to shearing of flux-tubes in the disk midplane, a state perhaps similar to magnetic frustrated convection \citep[e.g.,][]{Pen:2003}.

\subsection{Angular momentum transport versus polar angle} \label{sec:polar}

\begin{figure}
    \includegraphics[width=\columnwidth]{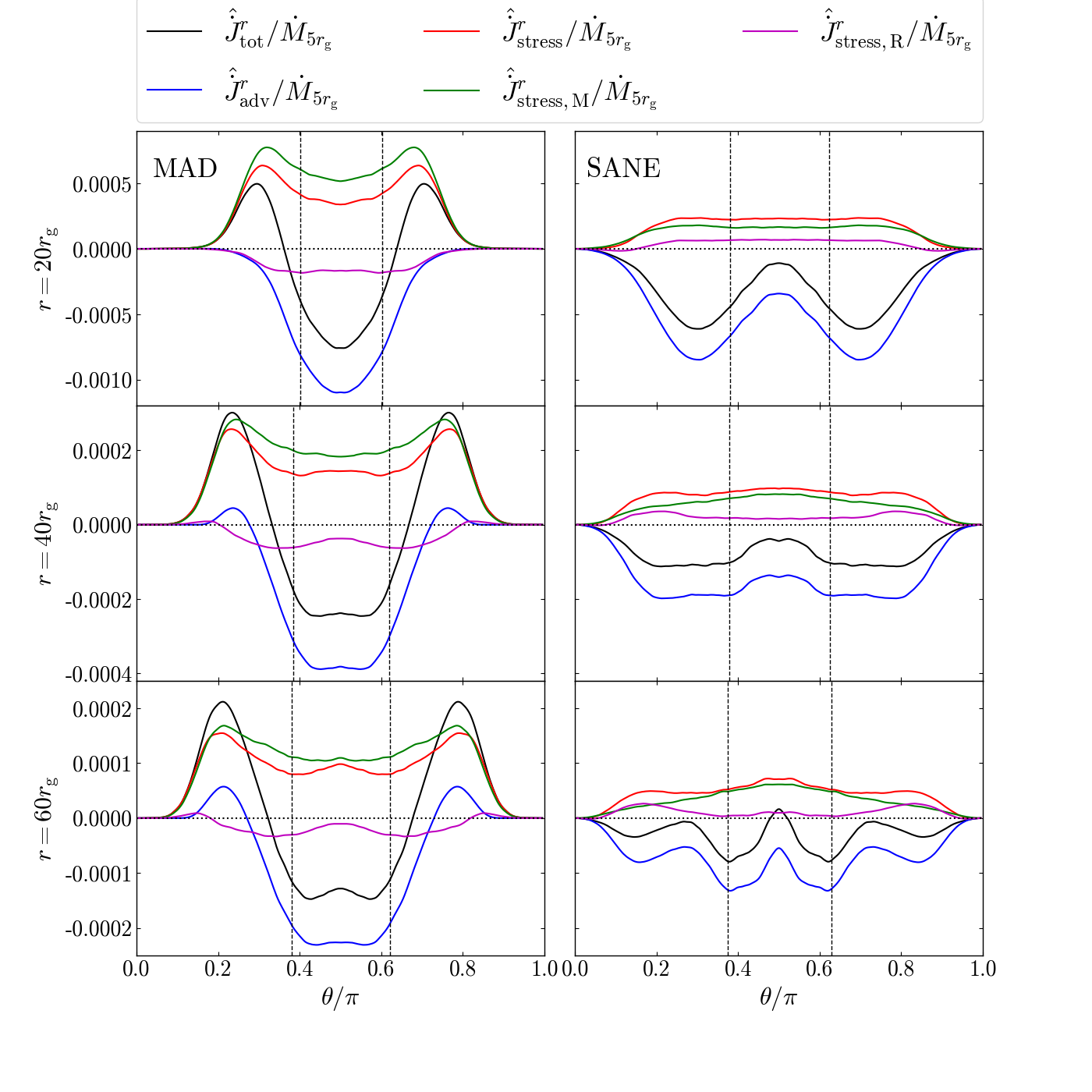}
    \caption{Here we show the angular dependence of the $\Mdot-$normalized angular momentum flux components at different radii: $20, 40, 60 \rg$, with the vertical lines indicating the disk scale aspect ratio at each radius. We see a clear outward flux in the wind region of the MAD model (left column), which is in contrast to the SANE model (right column) where the fluxes are at their maximum in the disk bulk.}
    \label{fig:theta_Jdot}
\end{figure}

\noindent Next we look at the variation of different components of the radial flux of the angular momentum over the polar coordinate $\theta$ at different radii across the disk and the wind. Figure~\ref{fig:theta_Jdot} shows the radial components of $\Jtot$, $\Jadv$, $\Jout$, $\JoutM$ and $\JoutR$, all normalized by the corresponding accretion rate, at $r=20, 40$ and $60\rg$. The absolute values of the different specific angular momentum flux components (i.e., $\dot{J}_{\rm X}/\dot{M}$) are larger in the MAD model, highlighting the importance of strong magnetic stresses in angular momentum transport. In the $\Jtot$ profiles of the MAD model, we see that there is a net outward flux (indicated by positive $\Jtot$ values) just outside of the disk at all 3 radii even though there is no persistently strong wind at $r=20\rg$. 

There seems to be a decrease in the magnitude of $\Jtot$ in the midplane of the SANE disk, which suggests that even though small-scale eddies dominate the angular momentum transport in this region, these features are washed out due to the $\theta-$symmetrization of $\Jtot$. This is why we see small values of $\Jtot$ in the SANE disk midplane for $r=60\rg$ as the disk is marginally in inflow equilibrium at this radius due to the presence of large scale eddies. Unlike the MAD model, $\Jtot$ is always negative (i.e., points inward) in the SANE model due to the absence of strong winds (also see Fig.~\ref{fig:Jdot_2D}). Thus, when we calculate the shell-integrated total flux $\dot{J}^r_{\rm int}$, the absolute value of this quantity is large. For the MAD model, as we integrate over $\theta$, the inward (i.e., negative) $\Jtot$ in the MAD disk cancels out with the outward $\Jtot$ in the wind, resulting in a small net flux (see Fig.~\ref{fig:Jdot_radial}). 

In the case of $\Jadv$, we see similar profiles as $\Jtot$ since the disk is advection-dominated in both models. If we consider the wind region in the MAD model, we see that the stress contribution to the outward angular momentum flux $\Jout$ is far larger than the advective contribution $\Jadv$. The angular momentum transferred by the bulk motion of the wind is far less important than the magnetic stress, clearly indicating a magnetically driven wind. In the SANE model, the MRI is the primary agent behind outward angular momentum transport and so, $\Jout$ is largest in the equatorial region. As we have noted before, the sign of $\JoutR$ is opposite in the two models, with Reynolds stresses bringing in angular momentum on average in the MAD disk and removing angular momentum in the SANE disk. In either case, the Reynolds component is smaller in magnitude when compared to the Maxwell component.

The angular momentum flux becomes negligible in the polar region (i.e., $\theta \sim 0$ and $\pi$). This is because there is hardly any gas or strong magnetic fields here. For spinning BHs, the jets occupy the polar region and exert an outward force expelling gas and carrying out angular momentum. Indeed, for spinning BHs in the MAD state, jets dominate outward angular momentum transport as discussed in Sec.~\ref{sec:spinup}.

\begin{figure}
	\includegraphics[width=0.8\columnwidth]{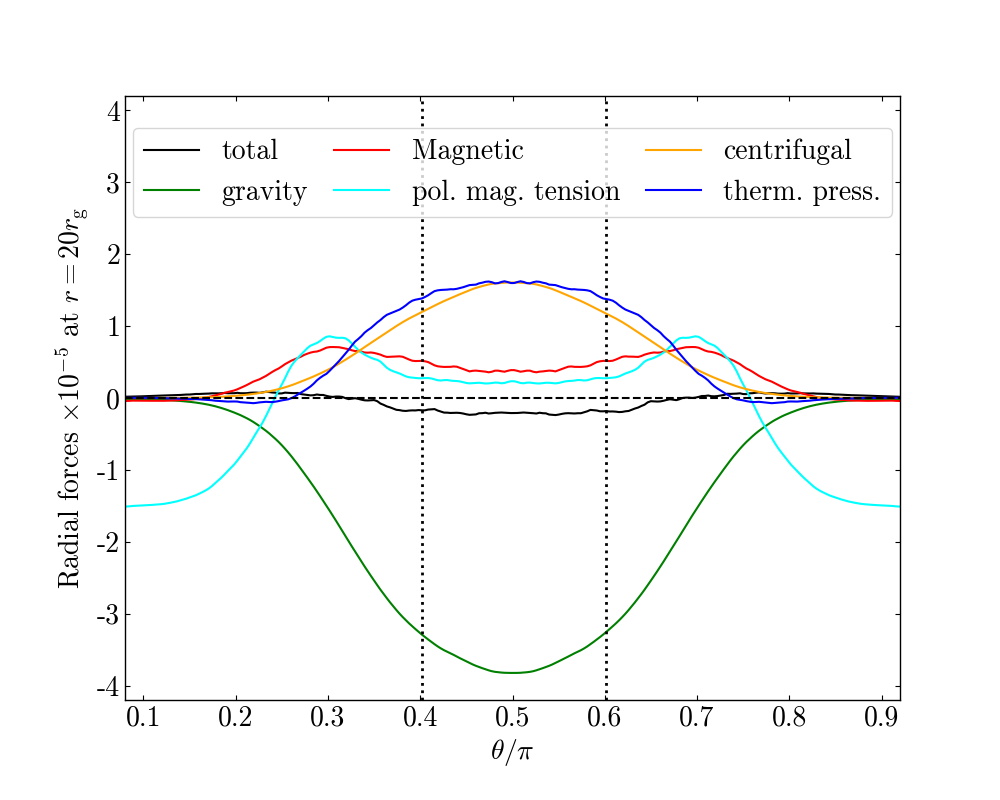}
	\includegraphics[width=0.8\columnwidth]{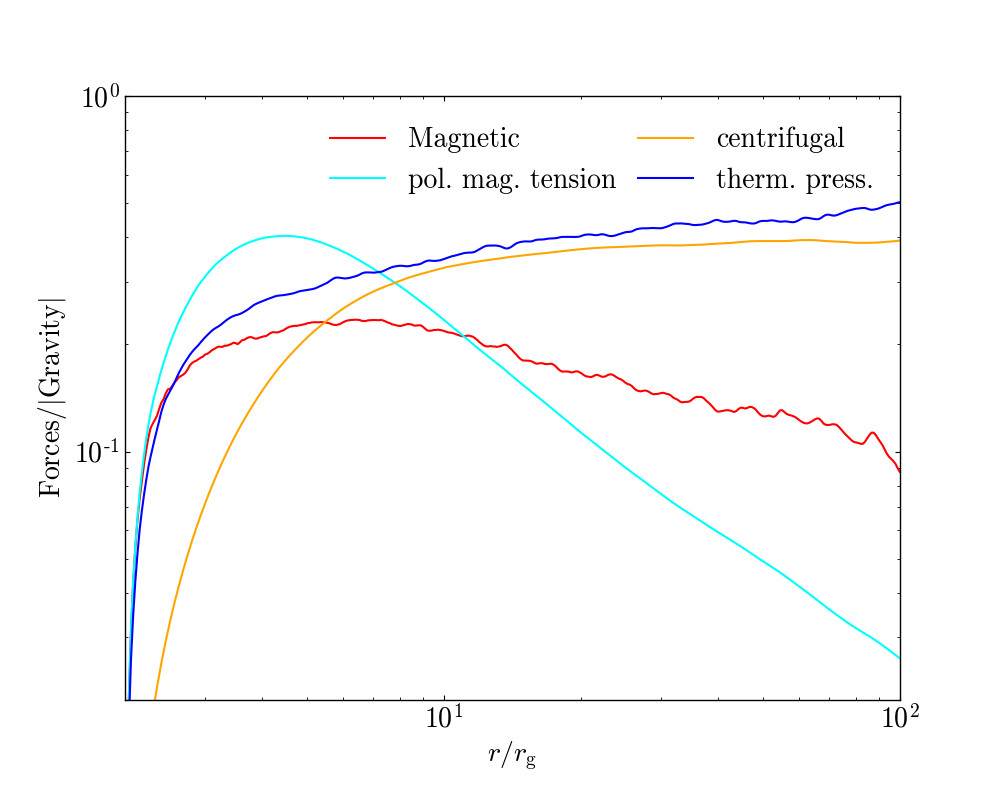}
    \caption{The centrifugal force and thermal pressure generally support the disk while magnetic forces dominate the wind. We show the different radial forces due to the thermal pressure, magnetic fields and gravity for the MAD model. Top: Radial forces at radius $r=20\rg$, calculated over the time period $2.4-2.9\times10^5\tg$. The vertical dotted lines show the disk scale aspect ratio $h/r$ at $20\rg$. Bottom: Radial profile of disk-averaged forces. We see that the poloidal magnetic tension dominates the total magnetic force in the disk at least for $r\lesssim20\rg$.}
    \label{fig:forces}
\end{figure}

\subsection{Force balance} \label{sec:forces}

As we have seen, magnetic fields play a leading role in regulating angular momentum balance in MADs. The question arises as to which component of the magnetic field, poloidal or toroidal, is responsible for accelerating gas in the wind in the MAD regime? Earlier analysis of MAD simulations defined the MAD regime out to a radius where the poloidal magnetic tension is able to balance gravity \citep[e.g.,][]{narayan03,mckinney12}. This notion was recently questioned by \citet{Begelman2022} who proposed that the toroidal field dominates instead and drives the saturation of the magnetic flux in the MAD regime. Figure~\ref{fig:pressure} shows that the pressure due to the radial field is much larger than that from the toroidal field within the inner $10\rg$, which is not the case in \citet{Begelman2022}. It is possible that the difference arises due to the lack of BH spin in our models as a spinning BH would twist vertical fields in the azimuthal direction and launch a jet. Then how important are the poloidal fields in driving mass loss by powering winds? 

To study the acceleration of gas in a steady-state MAD, we calculate the radial forces due to the different components of the magnetic field using the conservation equation $T_{\nu;\mu}^\mu = 0$ where the energy-momentum tensor $T^\mu_{\nu}$ is given by
\begin{equation}
    T^{\mu}_{\nu} = \left(\rho + u_{\rm gas} + p_{\rm gas} + b^2\right )u^\mu u_\nu + \left(p_{\rm gas} + \frac{b^2}{2}\right)\delta_{\nu}^{\mu} - b^{\mu} b_{\nu}.
\end{equation}
Assuming steady state, axisymmetry and a Schwarzschild metric, the radial momentum equation reduces to

\begin{eqnarray}
T^{\mu}_{r;\mu} &=& \frac{1}{\sqrt{-g}}\partial_{r}(\sqrt{-g}T^{r}_r) + \frac{1}{\sqrt{-g}}\partial_{\theta}(\sqrt{-g}T^{\theta}_r) \nonumber \\ 
  &-& \frac{1}{r}(T^{\theta}_{\theta}+T^{\varphi}_{\varphi}) + \Gamma^{t}_{r t}(T^{r}_{r} - T^{t}_{t}) = 0.
\label{eq:force}
\end{eqnarray}

\noindent First, we note that gravity, which is described by the mass parameter $M$, appears only in the Christoffel symbol $\Gamma$ in the last term of eq.~\eqref{eq:force}. We identify this with gravity:
\begin{eqnarray}
    \qquad\qquad\qquad {\rm gravity}&=&\Gamma^{t}_{r t}(T^{r}_{r} - T^{t}_{t}) \nonumber \\
    &=&\frac{M}{r^2\left(1-\frac{2M}{r}\right)}\left(T_r^r-T_t^t\right).
\end{eqnarray}

\noindent For the remaining terms in eq.~\eqref{eq:force}, we split $T_\nu^\mu$ into separate contributions corresponding to energy density, gas pressure and magnetic stress:
\begin{eqnarray}
{\rm energy~density}&:& {}^e T_\nu^\mu = \left(\rho + u +\frac{b^2}{2}\right)u^\mu u_\nu , \\ 
{\rm gas~pressure}&:& {}^p T_\nu^\mu = p u^\mu u_\nu + p \delta_\nu^\mu, \\
{\rm magnetic~stress}&:& {}^mT_\nu^\mu = \frac{b^2}{2} u^\mu u_\nu + \frac{b^2}{2} \delta_\nu^\mu -b^\mu b_\nu.
\end{eqnarray}

\noindent The energy density part of $T_\nu^\mu$ behaves like inertia and we treat its contribution as the dynamical part of the radial momentum equation:
\begin{eqnarray}
{\rm dynamical} &:=& \frac{1}{\sqrt{-g}} \frac{\partial}{\partial r}\left(\sqrt{-g}\,{}^eT_r^r\right) + \frac{1}{\sqrt{-g}} \frac{\partial}{\partial \theta}\left(\sqrt{-g}\,{}^eT_r^\theta\right) \nonumber \\ &~&~~~~-\frac{1}{r}\left({}^eT_\theta^\theta + {}^eT_\phi^\phi\right).
\label{eq:dyn}
\end{eqnarray}
\noindent Similarly the pressure term is
\begin{eqnarray}
{\rm pressure} &:=& \frac{1}{\sqrt{-g}} \frac{\partial}{\partial r}\left(\sqrt{-g}\,{}^pT_r^r\right) + \frac{1}{\sqrt{-g}} \frac{\partial}{\partial \theta}\left(\sqrt{-g}\,{}^pT_r^\theta\right) \nonumber \\ &~&~~~~-\frac{1}{r}\left({}^pT_\theta^\theta + {}^pT_\phi^\phi\right).
\label{eq:pres}
\end{eqnarray}

\noindent Finally, in the case of the magnetic stress contribution, we focus on the outward poloidal magnetic tension:
\begin{equation}
{\rm pol.~mag.~tension} := \frac{1}{\sqrt{-g}} \frac{\partial}{\partial \theta}\left(\sqrt{-g}\,{}^mT_r^\theta\right).
\label{eq:polmag}
\end{equation}

We first axisymmetrize and time-average each term within brackets in eq.~\eqref{eq:dyn}, \eqref{eq:pres} and \eqref{eq:polmag} over the final $5\times10^4\tg$ for the MAD model. Then we symmetrize these terms in $\theta$ across the equatorial plane, taking care that the $T^{\theta}_{r}$ terms are anti-symmetric in $\theta$. This is the same averaging process we used to construct the angular momentum flux components. Once we have the $t\varphi-$averaged and $\theta-$symmetrized versions of the bracketed terms, we perform the differentials in $r$ and $\theta$ on the terms with $T^r_r$ and $T^{\theta}_{r}$ respectively as given in the equations. To get the disk-averaged quantities, we average over $\theta$ using gas density as a weighting term. 

Figure~\ref{fig:forces} shows the polar structure of the radial forces at $r=20\rg$ (upper panel) and the disk-averaged radial profile (lower panel). We see that the disk, indicated by $h/r$ (vertical dotted lines), is mainly supported by rotation and thermal pressure against gravity. The wind, on the other hand, is marginally dominated by forces due to the magnetic field, hence showing that magnetic fields indeed are instrumental in driving winds in MADs. The magnitude of the magnetic pressure gradient and the hoop stress, both of which prominently feature the toroidal field component, are larger than the poloidal magnetic tension. However, they are opposite in sign and roughly cancel each other within $\sim 20\rg$, suggesting that the toroidal field does not play a significant role in the radial force balance in the inner disk \citep[also noted by][]{Begelman2022}. Instead it is the poloidal magnetic tension that peaks in the wind region and is thereby responsible for launching winds from the inner disk.

Forces due to the magnetic field dominate within $r\sim 7\rg$ in the MAD model and support the disk against gravity. Thermal pressure and centrifugal forces support the disk at larger radii ($r\gtrsim 20\rg$). This suggests that the force balance equation used to traditionally define the ``MAD'' regime \citep[][]{narayan03}, i.e., equating the gravity and the poloidal magnetic tension terms, might only work in the inner few gravitational radii, where very strong vertical fields are present. 

While not shown here, the centrifugal force (and thermal pressure to a lesser extent) primarily supports the SANE disk against gravity at all radii within $r\sim 100\rg$. The poloidal magnetic tension is, even at its largest, an order of magnitude smaller than the centrifugal force. This is the reason why winds in the MAD model are more powerful than those in the SANE model, and therefore more efficient in removing angular momentum. 

\begin{figure*}
    \includegraphics[width=\textwidth, trim={2cm 2cm 2cm 1cm},clip]{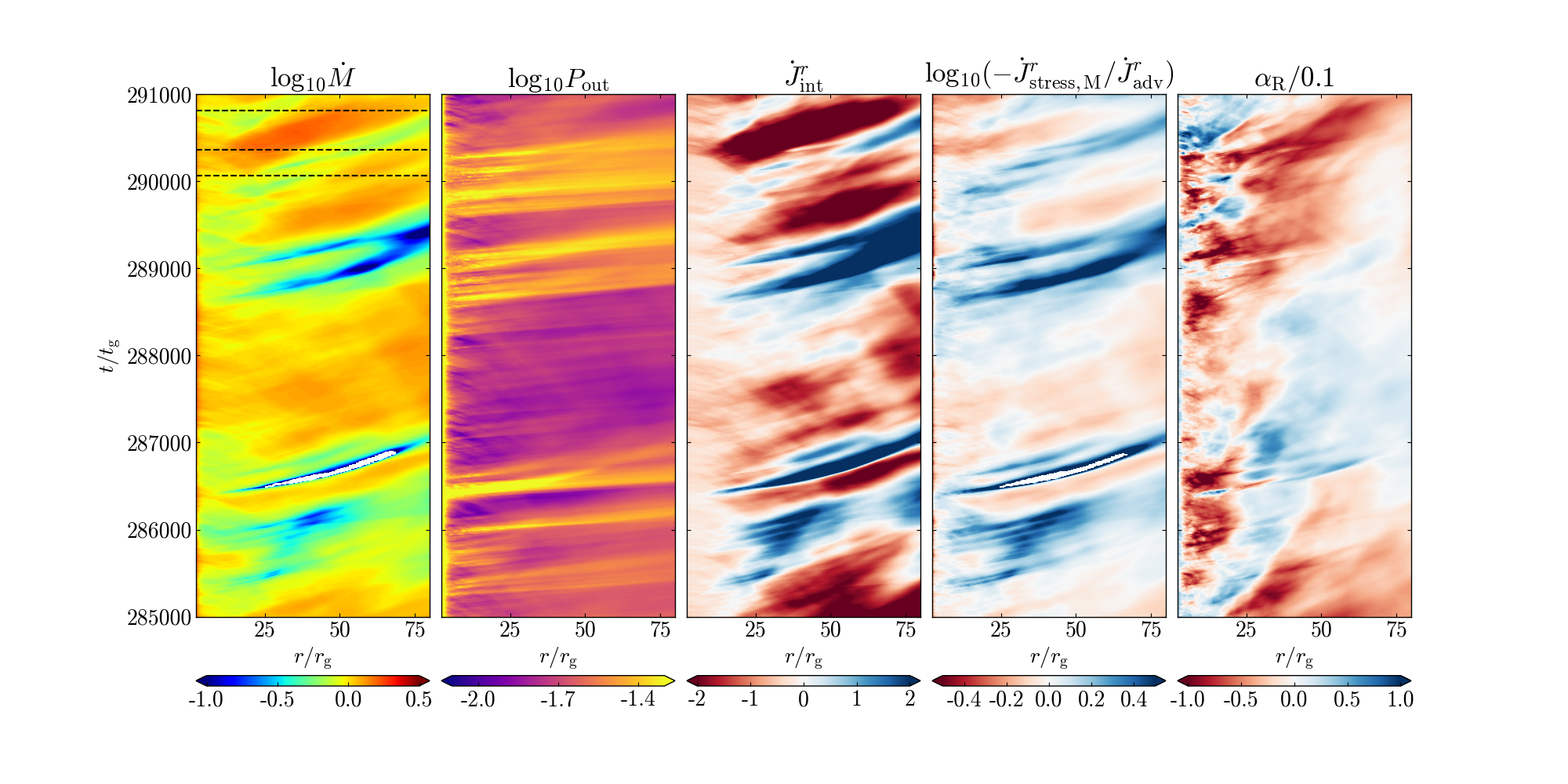}
    \caption{Changes in the accretion rate ($\dot{M}$), outflow power ($P_{\rm out}$), total angular momentum flux ($\dot{J}^r_{\rm int}$), the ratio of the shell-integrated Maxwell and advective fluxes ($\dot{J}^r_{\rm stress, M}/\dot{J}^r_{\rm adv}$), and Reynolds stress ($\alphaR$) are correlated in MADs. We show the time-radial plots for each quantity for the MAD model. The white regions in $\Mdot$ indicate a net outward mass flux. The dashed lines in the $\Mdot$ panel indicate time snapshots over which we track one magnetic flux eruption event in Fig.~\ref{fig:flux_eruption2D}. $\dot{J}^r_{\rm adv}$ is generally directed inwards towards the BH, and so we use a negative sign for the ratio calculation. White regions in the $\dot{J}^r_{\rm stress, M}/\dot{J}^r_{\rm adv}$ plot indicate a net outward angular momentum flux due to advection.}
    \label{fig:rt_plot_2D}
\end{figure*}

\begin{figure*}
    \includegraphics[width=\textwidth]{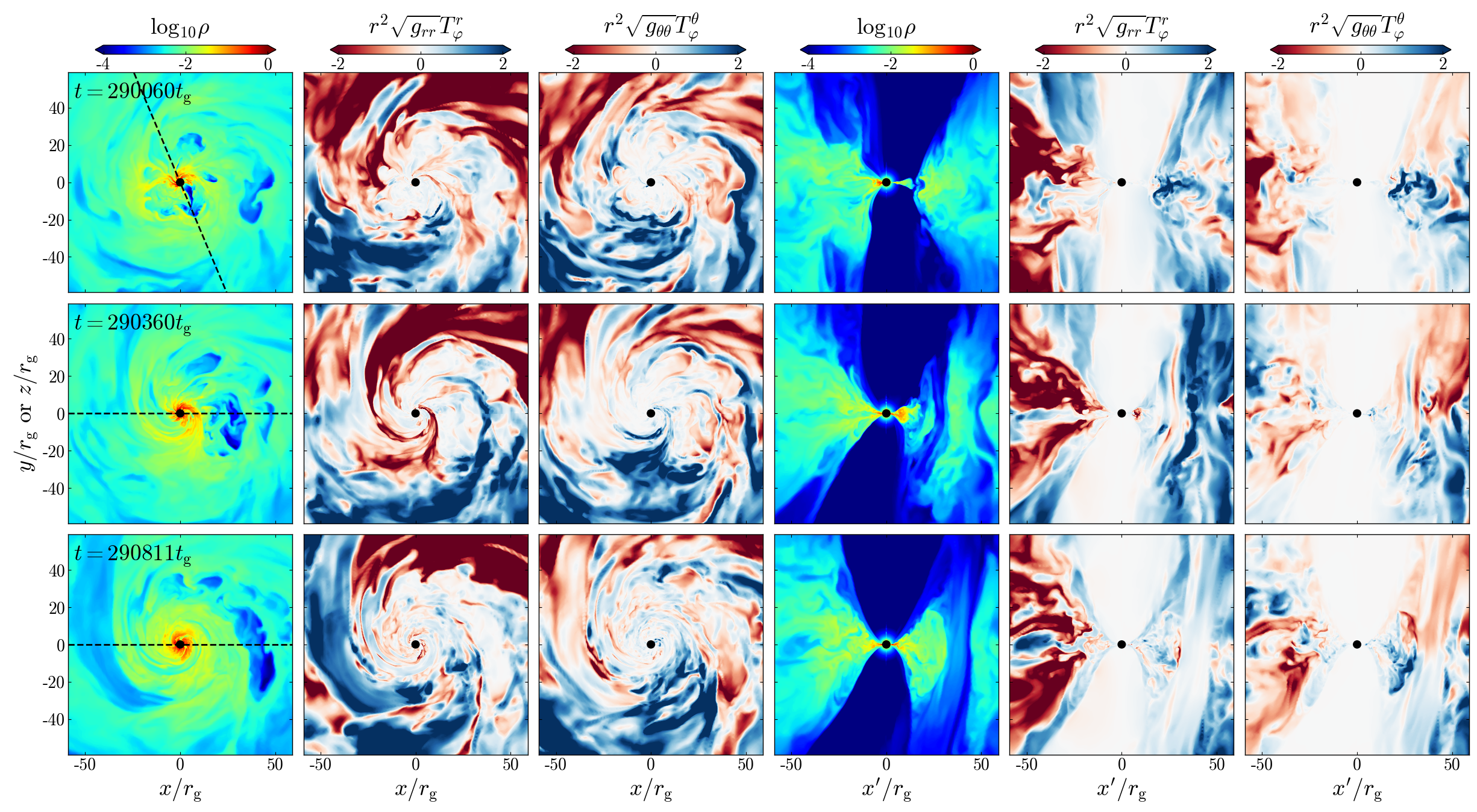}
    \caption{Magnetic flux eruptions trigger strong disk winds and drive outward angular momentum transport in MADs. We show midplane and vertical cross-sections of gas density $\rho$ and the radial and polar fluxes of the angular momentum at 3 different times corresponding to the dashed lines in Fig.~\ref{fig:rt_plot_2D}, tracking the evolution of a particular flux eruption. The dashed lines in the first column indicate the azimuthal angle at which we take the corresponding vertical slices. The blue (red) regions indicate outward (inward) fluxes for $T^r_{\varphi}$ and clockwise (counterclockwise) fluxes for $T^{\theta}_{\varphi}$. Flux eruptions push gas outward, creating outward fluxes as seen from the right halves of the vertical slices. The left halves of these plots show a region of the disk that does not have a flux eruption. Hence, turbulence drives a net inward transport of angular momentum, similar to a SANE disk.}
    \label{fig:flux_eruption2D}
\end{figure*}

\section{The transient nature of the MAD state} \label{sec:MADtime}

The time evolution plot (Fig.~\ref{fig:time}) in Sec.~\ref{sec:time} clearly shows that the MAD angular momentum flux is highly variable as compared to the SANE case. Indeed the ratio of the standard deviation ($\sigma$) and the mean ($\mu$) of $\dot{J}^r_{\rm int}/\Mdot$, calculated at $5\rg$ over $2.4-2.9\times10^5\,\tg$ gives $\sigma/\mu=0.135$ and $0.47$ for the SANE and MAD models respectively, despite the $\sigma/\mu$ values of the accretion rate of both models being roughly similar ($\sim 0.26$ and $0.3$ for SANE and MAD). Further, as we noted earlier, the time-averaged viscosity due to Reynolds stress in the MAD model has a sign opposite to that in the SANE model. Does the behavior of the Reynolds stress in MADs also change with time?

Figure~\ref{fig:rt_plot_2D} shows the time-variable nature of the radial profiles of the accretion rate $\Mdot$, the outflow power $P_{\rm out}$, the shell-integrated radial flux of the total angular momentum $\dot{J}^r_{\rm int}$, the ratio of the shell-integrated Maxwell and advection components of the angular momentum flux ($\dot{J}^r_{\rm stress, M}/\dot{J}^r_{\rm adv}$), and the disk-averaged Reynolds stress $\alphaR$ of the MAD disk. Over our selected time segment of $2.85-2.91\times10^5\tg$, we see two strong magnetic flux eruption events around $286500\tg$ and $289000\tg$ when multiple eruptions occurring at different azimuths around the BH push $\Mdot$ to near-negative values (i.e., a net outward mass flux). During these eruptions, magnetic flux-tubes propagate outward into the disk and push against the inflowing gas, thus increasing the outflow power and thereby launching winds. 

For the angular momentum flux, we see that the pattern in $\dot{J}^r_{\rm int}$ matches that of $\Mdot$ and $P_{\rm out}$, with the angular momentum flux changing from inward to outward net flux during magnetic flux eruption events. During eruptions, as vertical fields get injected into the disk, we see that the Maxwell stress component of the outward angular momentum flux dominates over the inward advection component (Fig.~\ref{fig:rt_plot_2D}, fourth panel). This results in a net outward $\dot{J}^r_{\rm int}$. Thus, this result establishes that there is a close link between eruptions, winds and outward angular momentum transport. It is interesting to note that since $\dot{J}^r_{\rm adv}$ behaves similar to $\Mdot$, strong eruption episodes can also result in a net outward advective momentum flux component, producing a significantly strong wind angular momentum flux.

Next we see that there is a change in the sign of the Reynolds stress $\alphaR$ during certain time periods extending over large portions of the inner disk. The pattern in $\alphaR$ is not an exact match to $\dot{J}^r_{\rm int}$ but this is expected since Maxwell stresses dominate $\Jout$. We note that between $287000-288500\tg$, even though the net angular momentum flux is negative (i.e., net inward flux), the absolute value is smaller as compared to $t=285100\tg$ or $290500\tg$ since $\JoutR$ is pointing outward. The eruption event at $t\sim286500\tg$ pushes out magnetic flux, causing the disk to be SANE-like beyond $20-40\rg$. Interestingly, this suggests that during this period, MRI in the disk bulk \citep[indicated by $\alphaR>0$;][]{Pessah:2006} may become strong enough to regulate angular momentum, hence causing the $\dot{J}^r_{\rm int}$ to be lower than average. With time, magnetic flux re-accumulates in the inner disk and we transition back into the MAD state. Such behavior is completely absent in the SANE model where MRI is the dominant mechanism of angular momentum transport and both $\alphaR$ and $\dot{J}^r_{\rm int}$ maintain the same sign throughout the disk at all times. The regeneration time for poloidal magnetic flux varies between a few hundred to a thousand $\tg$ \citep[also see][]{Ripperda2022}, which results in multiple $\alphaR=0$ regions often seen in time-averaged plots of $\alphaR$ \citep[e.g., Fig. 5 of][]{liska_tor_2019}.

So far we have established that flux eruptions, winds and outward angular momentum flux are strongly correlated in MADs (also see Sec.~\ref{sec:polar}). How does the whole picture of angular momentum transport in non-spinning BH MADs then fit together? Figure~\ref{fig:flux_eruption2D} shows the midplane and vertical cross-sections of the MAD model at three different times (indicated by the dashed lines in the $\Mdot$ plot of Fig.~\ref{fig:rt_plot_2D}). For the radial flux $T^r_{\varphi}$, the color scheme indicates outward radial fluxes (i.e., positive values) in blue and inward fluxes in red. In the case of $T^{\theta}_{\varphi}$, clockwise and counterclockwise fluxes in the $\theta$ direction are shown in blue and red. We track one specific magnetic flux-tube as it travels outward through the disk and experiences shearing due to the rotating gas. 

As the magnetic flux-tube pushes out against the accreting gas, it triggers outward movement of angular momentum (indicated by the blue region in both the midplane and vertical plots of $T^r_{\varphi}$). The outward motion of the flux-tube injects strong vertical fields into the disk, reinvigorating winds and transporting angular momentum vertically, i.e., we get a counterclockwise $T^{\theta}_{\varphi}$ flux in the upper hemisphere of the disk and a clockwise flux in the lower hemisphere. It is particularly noteworthy that the strength of $T^{\theta}_{\varphi}$ is on par with $T^{r}_{\varphi}$, especially near the disk midplane, highlighting the strong vertical nature of angular momentum transport during a flux eruption event. Finally, the flux-tube dissipates in the disk where the azimuthal shearing is the strongest and disk angular momentum flux returns to pre-eruption levels. The loss in angular momentum is strongest during the time periods when we see multiple magnetic flux events. Indeed, during such times, the magnetic flux within the inner $\sim 10\rg$ decays to sub-MAD levels and we see similar properties as a SANE accretion flow, such as a positive $\alphaR$.  

It is interesting to compare the left and right-hand sides of the vertical cross-section plots. The right-hand side captures the effect of a magnetic flux eruption on the flow of angular momentum in the disk, showing a well structured outward angular momentum flux. On the other hand, the left side shows a region that is not undergoing a flux eruption. This region exhibits a mix of inward and outward fluxes for both of the radial and the polar components, indicating a turbulent flow more typical of a SANE disk. Indeed, we see that inward radial fluxes of the angular momentum dominate over a large portion of the MAD disk, with a weak outward flux in the wind region, similar to the SANE model. The distinct contrast between the two sides of the accretion flow in the same MAD model lends further support to the highly variable and non-axisymmetric nature of the MAD state where MRI may be suppressed in only a part of the disk depending on the presence of strong vertical fields.

\section{Discussion} \label{sec:discussion}

\subsection{Gas distribution around Sgr A$^*$ and M87$^*$} \label{sec:density_slope}

\begin{figure}
	\includegraphics[width=\columnwidth]{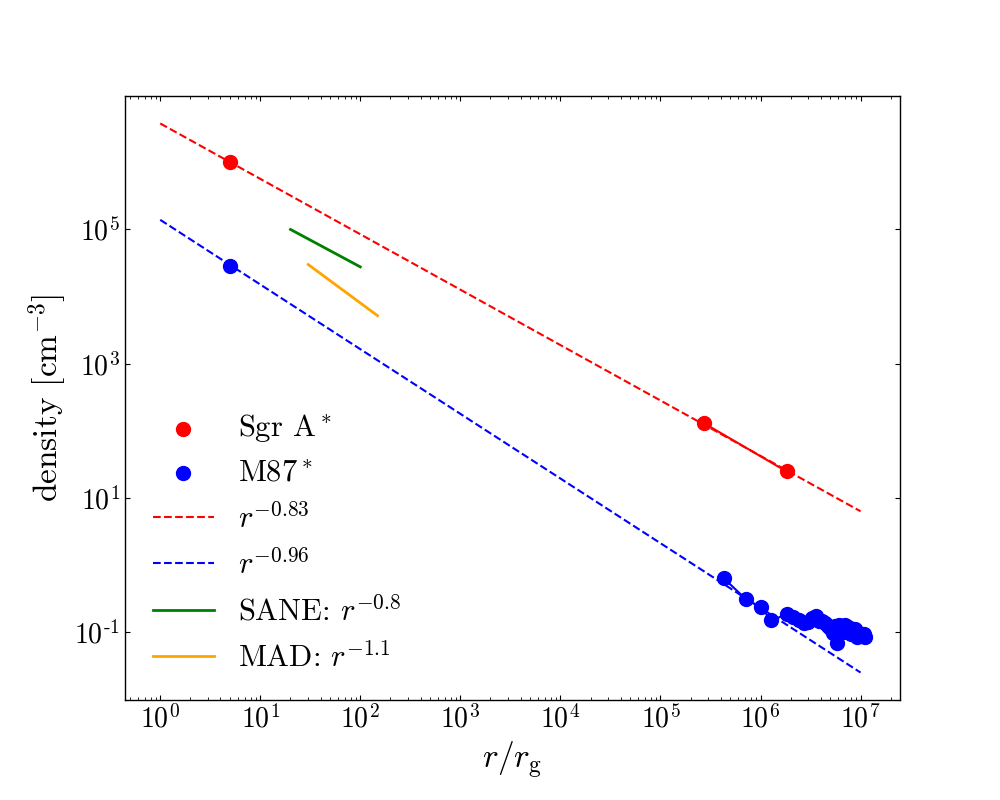}
    \caption{We compare the gas density profile of the accretion flow in Sgr A$^*$ and M87$^*$ with the MAD and SANE models (from Fig.~\ref{fig:disk_radial_time}). The observational data and the simulations show consistent radial profiles: $\sim \rho \propto r^{-1}$. We take the horizon-scale densities from \citet{EHT_M87_2019_PaperV, EHT_SgrA_2022_PaperV} and the Bondi-scale data from Fig.~6 of \citet{Alexander:2016}, originally from \citet{Baganoff:03} for Sgr A$^*$ and \citet{Russell:2015_M87} for M87$^*$. The dashed lines indicate power-law fits connecting the near-horizon data points with the Bondi-scale data.}
    \label{fig:density_slope}
\end{figure}

One of the fundamental questions about supermassive BH accretion is how gas is distributed over multiple length scales. The BH's gravitational field broadly dictates gas dynamics from near the event horizon out to almost the Bondi radius ($\sim 10^{5-6}r_{\rm g}$). Thus, we expect the gas to maintain a coherent power-law profile over roughly 6 orders of magnitude in distance, beyond which the large-scale turbulent structures in the interstellar medium become prominent. The Event Horizon Telescope (EHT) results on M87$^*$ \citep[][]{EHT_M87_2019_PaperV} and Sgr A$^*$ \citep[][]{EHT_SgrA_2022_PaperV} provide crucial information about the gas density distribution near the BH, and enable us to connect the event horizon and Bondi radius scales. 

Figure.~\ref{fig:density_slope} shows the radial profiles of the gas density $\rho$ as inferred from sub-millimeter and X-ray observations of Sgr A$^*$ \citep[][]{EHT_SgrA_2022_PaperV, Baganoff:03} and M87$^*$ \citep[][]{EHT_M87_2019_PaperV, Russell:2015_M87}. Assuming a one-zone uniform sphere of radius $5r_{\rm g}$, plasma-$\beta \sim 1$ and optically thin thermal synchrotron emission, the EHT estimates for the gas density in M87$^*$ and Sgr A$^*$ are $2.9\times 10^{4} ~{\rm cm}^{-3}$ and $10^{6} ~{\rm cm}^{-3}$, respectively. Fitting for the observationally-inferred gas density data points, we get slopes of $-0.83$ and $-0.96$ for Sgr A$^*$ and M87$^*$, respectively. For M87$^*$, we see a transition to a flatter power-law profile for $r\gtrsim10^6\rg$ as the interstellar medium begins to dominate beyond the Bondi radius. This distance of $\sim10^6\rg$ in M87$^*$ also roughly corresponds to the position of the HST-1 knot and coincides with a transition in the jet shape from a parabolic to a conical profile \citep[][]{asadanak2012}. It is possible that the change in the density slope causes the jet to over-collimate, which results in a knotted feature and a conical outflow.

The density profiles from the observations are consistent with the radial slopes from the MAD ($\rho\propto r^{-1.1}$) and SANE ($\rho\propto r^{-0.8}$) models (solid lines in Fig.~\ref{fig:density_slope}). For the slope calculation from the simulations, we time-average the density profiles over $(2.4-2.9)\times 10^5\tg$. The top row in Fig.~\ref{fig:disk_radial_time} shows that the density profiles within $r\lesssim100\rg$ in the MAD and SANE models are flatter at early times and only converge toward a slope of $-1$ near the end of the simulation runtime. This occurs because the accretion flow within $100\rg$ only reaches inflow-outflow equilibrium at $t\gtrsim2\times 10^5\tg$ (see Sec~\ref{sec:radial}). Though evolved over a shorter dynamical time, the GRMHD simulations of accretion onto Sgr A$^*$ performed by \citet{Ressler:2020:sgra_MAD} also exhibit a radial slope of $\sim-1$ for the gas density, matching well with our simulations. 

\begin{figure}
	\includegraphics[width=\columnwidth]{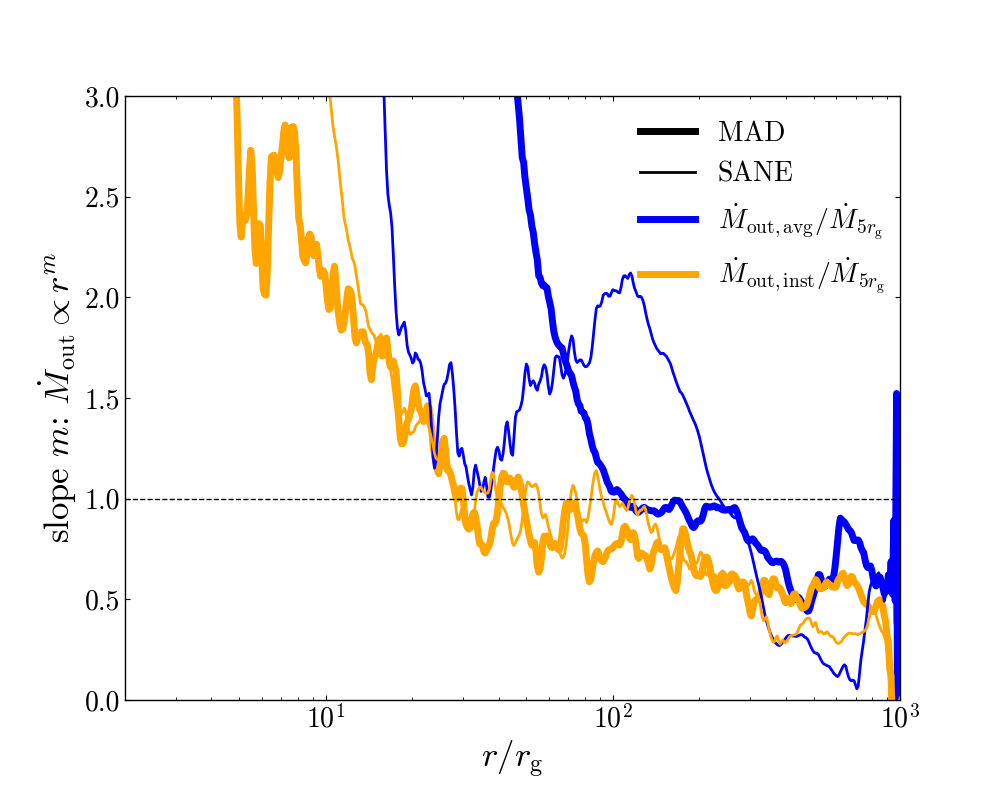}
    \caption{We calculate the radial slope of the mass outflow rates $\Mdot_{\rm out}\propto r^m$ for both the average and instantaneous mass outflow rates (see Fig.~\ref{fig:Mdot_radial}). The slope $m$ is used to determine the radial density profile according to the ADIOS model and is roughly $1$ around the inflow equilibrium radius ($r\sim100\rg$). However, we caution that $\Mdot_{\rm out}$ drop rapidly as we approach the BH, and therefore, the slope in this region may not much relevance.}
    \label{fig:Mdotout_slope}
\end{figure}

The radial density profile is flatter than that expected for a spherically symmetric Bondi accretion flow or a classic (non-wind) ADAF ($\rho \propto r^{-3/2}$). This suggests that outflows may indeed be important for regulating the radial gas distribution even though our simulations indicate that the mass outflow rates are small within $r\lesssim100\rg$ (see Fig.~\ref{fig:Mdot_radial}). In the advection-dominated inflow-outflow solution \citep[ADIOS;][]{Blandford_begelman_1999}, the predicted density scales as $\rho\propto r^{-3/2+m}$, where the mass outflow rate $\dot{M}_{\rm out}\propto r^m$. The results shown in Fig.~\ref{fig:density_slope} require $m\approx 0.5$ and $0.7$ for M87$^*$ and Sgr A$^*$. The radial profiles of the average and instantaneous mass outflow rates for the MAD and SANE models become flatter with larger radius, with the radial slope roughly between $\sim0.7$ to $1$ at approximately $100\rg$ (Fig.~\ref{fig:Mdotout_slope}). The mass outflow rate from \citet{Ressler:2020:sgra_MAD} shows a radial slope $\sim1$, which is in rough agreement with our results. However, since $r=100\rg$ is near the outer edge of the inflow equilibrium radius in our simulations, we require longer simulation runtimes to confirm whether the slope indeed decreases to 0.5. It is also possible that for jetted BHs such as M87$^*$, jet-wind interactions might change the density profiles that we find here. We leave the study of density profiles from jetted BHs as future work.

\subsection{Black hole spinup} \label{sec:spinup}

\begin{figure}
	\includegraphics[width=\columnwidth]{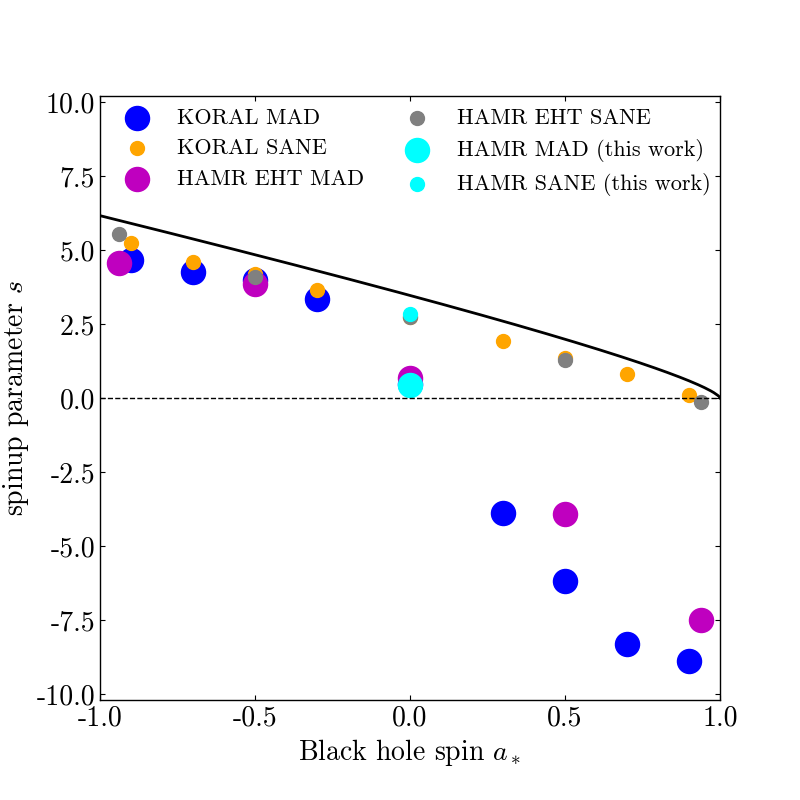}
    \caption{SANE accretion flows spinup BHs while the MAD state reduces the BH spin magnitude over long timescales. We show the spinup parameter $s$ (eq.~\ref{eq:spinup}) for the MAD and SANE models (cyan circles) in the present work. For comparison, we show $s$ for KORAL simulations of MAD \citep[black circles;][]{Narayan2022} and SANE disks (orange circles) with nine different BH spins, and \hamr{} MAD (magenta circles) and SANE (gray circles) simulations from \citet{EHT_SgrA_2022_PaperV}. The black solid line indicates the spinup parameter for a standard thin disk \citep[][]{Shapiro:2005}.}
    \label{fig:spinup}
\end{figure}

The zero spin BH MAD and SANE models exhibit a net inward angular momentum flux. Thus, we expect the corresponding BHs to gain angular momentum over time. We can quantify the spinup via the following spinup parameter $s$ \citep{Shapiro:2005, Narayan2022}:
\begin{equation}
    s= \frac{da_*}{dt}\frac{M_{\rm BH}}{\Mdot} =-\frac{\dot{J}^r_{\rm int}}{\Mdot} - \frac{2\dot{E}a_*}{\Mdot},
    \label{eq:spinup}
\end{equation}
\noindent where $a_*$ is the BH spin parameter and $M_{\rm BH}$ is the BH mass. 

In the current work, we have only considered $a_*=0$ MAD and SANE models. MADs with spinning BHs exhibit extremely efficient jets with an energy output which can at times exceed the input accretion energy \citep[][]{tch11}. Thus, jets in MADs can significantly affect the BH's spin evolution. For the discussion in this subsection, we also include previous results from MAD simulations described in \citet{Narayan2022}, who considered 9 different BH spins: $0.0$, $\pm 0.3$, $\pm 0.5$, $\pm 0.7$ and $\pm 0.9$, and calculated an average value of $s$ for each model over the time period $(5 - 10) \times 10^4 \tg$. Those simulations were performed using the GRMHD code KORAL \citep[][]{Sadowski:13_KORAL,Sadowski:14} and were each run for a duration of $10^5 \tg$. In addition, we include results from an equivalent set of SANE simulations with the same 9 spin values. These latter simulations employed the same basic setup as the MAD simulations in \citet{Narayan2022}, except that the initial magnetic field configuration was a set of quadrupolar poloidal field loops instead of a single dipolar loop, which ensured that the accretion flows remained SANE until the end of the simulation ($t = 3 \times 10^4 \tg$). Average $s$ values are calculated over the time range $(1.5-3)\times 10^4 \tg$. We also include \hamr{} SANE and MAD simulations from \citet{EHT_SgrA_2022_PaperV}, which considered 5 different BH spins: $0.0$, $\pm 0.5$ and $\pm 0.94$. These simulations were evolved to $\sim3.5\times 10^4\tg$. We calculate the spinup parameter for each model over the final $15000\tg$.

Figure~\ref{fig:spinup} shows the spinup parameter for KORAL and \hamr{} EHT simulations along with the zero spin MAD and SANE models considered in this work. First we note that the spinup values for the three simulation sets match very well. This shows that these spinup values are robust across different GRMHD codes, initial conditions and grid resolutions. There is a discrepancy in the high spin prograde MADs, probably due to the difference in density floors employed by the two codes. 

We see that the SANE models always exhibit positive spinup rates, similar to standard thin accretion disks \citep[][]{Shapiro:2005}. Hence, for SANE accretion flows, retrograde BHs spin down while prograde BHs spinup. The spinup-spindown equilibrium BH spin value for SANE disks is $a_{*, \rm eq}\simeq0.9$ and is slightly smaller than that for standard thin disks \citep[$a_{*, \rm eq}=0.998$;][]{Thorne:1974}. This value of $a_{*, \rm eq}$ is also consistent with early 2D SANE models \citep[e.g., $a_{*, \rm eq}\approx0.93$ from][]{Gammie:2004_bhspin}. 

For our zero spin SANE model, we find that $s=-\dot{J}^r_{\rm int}/\Mdot=2.85$ when time-averaged over $15,000-30,000\tg$, which is consistent with the values found from the corresponding KORAL/\hamr{} EHT simulation. The spinup parameter in the SANE model described in this paper decreases monotonically over time to $s=1.69$ when time-averaged over $2.4-2.9\times 10^5\tg$. The secular decrease in $s$ may possibly be due to the increase of the dimensionless horizon magnetic flux $\phi$ at $t\gtrsim 10^5\tg$ (see Fig.~\ref{fig:time}). As noted in Sec.~\ref{sec:time}, at this time, the disk begins to lose axisymmetry and there is polar infall of gas, leading to an increase in $\phi$. Further investigation of SANE simulations that exhibit $\phi$ values smaller than 5 over a long time period is required to check if $s$ indeed decreases over time.

The spinup parameter for the zero spin MAD model is $s=0.45$ (time-averaged over $2.4-2.9\times 10^5\tg$), which is a factor of a few smaller than the corresponding $s$ values for the thin accretion disk and the SANE model. Unlike the SANE model, the MAD model converges to $s\approx 0.45-0.55$ for $t\gtrsim$ a few $\times 10^4\tg$ and is consistent with the value obtained from the corresponding KORAL/\hamr{} EHT MAD simulation. For MADs with spinning BHs, jets dominate the angular momentum transport near the event horizon, effectively causing BH spindown \citep[][]{tch12proc,Narayan2022}.  From Fig.~\ref{fig:spinup}, we see that the magnitude of the BH spin would decrease over cosmological timescales except for very small values of prograde spin. For MADs, \citet{Narayan2022} and \citet{tch12proc} found an equilibrium spin value of $a_{*, \rm eq}\approx0.035$ and $0.07$ respectively. Thus, the MAD state is highly important for BH spin evolution, such as for long term sub-Eddington accretion in maintenance mode supermassive BHs \citep[e.g.,][]{Narayan2022} and super-Eddington accretion during gamma-ray bursts and tidal disruption events \citep[e.g.,][Curd, B., in prep]{Nathanail:2015}.

\section{Conclusions} \label{sec:conclusion}

\noindent In this work, we simulate two GRMHD accreting Schwarzschild black hole (BH) models, one with a weakly magnetized disk (i.e., standard and normal evolution, or SANE) and the other with a magnetically arrested disk (MAD), with high grid resolutions and for a duration up to $\sim 3\times10^5 GM_{\rm BH}/c^3$. Our primary goal is to investigate how mass loss and angular momentum transport take place in MADs, and our focus is on the role of disk physics and disk winds. Therefore, to avoid confusion from effects related to frame-dragging and the driving of relativistic jets, we limit our work to a non-spinning BH. In addition, by evolving the disk over a very long timescale, our models reach convergence in disk properties out to at least $100 GM_{\rm BH}/c^2$. The main results are as follows:
\begin{enumerate}
    \item The MAD state is a fundamentally transient condition as the horizon magnetic flux exhibits oscillatory behavior, rising to values above the average saturation point, and then decaying to a weak-field state due to the emergence of a magnetic flux eruption from near the BH event horizon. Thus we suggest that flux eruptions are a distinguishing feature of the MAD state in accretion disks in general.
    \item Absent relativistic jets, magnetic flux eruptions are the primary mechanism via which angular momentum is transported primarily vertically outwards in MADs, whereas the magnetorotational instability transports angular momentum outwards equatorially through the disk in the SANE model. The eruptions also strengthen the disk winds (up to outflow efficiencies of $5-10\%$) temporarily and initiate mass loss from the MAD disk. While the average mass outflow rate is only $60-80\%$ of the net accretion rate near the BH for both SANE and MAD models, the instantaneous mass outflow rate can become larger than the net accretion rate at large radii. The true mass loss rate via winds should lie between these two limits \citep[also see][]{Yuan:15}.
    \item On average, the Reynolds stress transports angular momentum inwards (towards the BH) in the MAD model, while the opposite is true in the SANE model. Further, the Reynolds stress changes direction frequently over time in the MAD model, suggesting that MRI might become prominent during certain time intervals.
    \item The poloidal magnetic tension dominates the net outward magnetic force on average, and provides support to the disk against gravity out to almost $10 GM_{\rm BH}/c^2$ \citep[in accordance with][]{narayan03}, suggesting that interchange instabilities due to poloidal fields regulate accretion within this radius. However, since the MAD state is highly transient and non-axisymmetric, MRI-driven accretion is also possible as suggested above. Additionally, it is difficult to state how far out the MAD state reaches in the disk. We speculate that the disk is saturated with magnetic flux out to at least $40-60 GM_{\rm BH}/c^2$, where the magnetic flux-tubes completely dissipate in the disk midplane due to azimuthal shearing.  
    \item The gas density scales as $\rho \propto r^{-0.8}$ in the SANE model and $\rho \propto r^{-1.1}$ in the MAD model. These slopes are consistent with the density profiles inferred for Sgr A$^*$ and M87$^*$. The slopes converge to the above values very late in the simulations, underscoring the importance of evolving these models to very long timescales.
    \item SANE accretion flows can potentially spin down retrograde BHs and spin up prograde BHs up to a spin of $\sim 0.9$. Jets from MAD accretion flows extract rotational energy from spinning BHs, and cause BH spindown in both retrograde and prograde systems \citep[e.g.,][]{Narayan2022}.  
    
\end{enumerate}

These results are in general agreement with previous work \citep[e.g.,][]{Narayan:2012, Begelman2022}. For spinning BHs, jets carry most of the angular momentum outwards, dragging the gas-rich disk winds to large velocities (nearly to relativistic levels) via gas mixing \citep[e.g.,][]{chatterjee2019}. When the magnetic flux reaches saturation in jetted BHs, the angular momentum loss is large enough to spindown the BH over time \citep[][]{tch12proc,Narayan2022}. However, even in jetted BHs, large-scale vertical magnetic fields in the winds still transport a significant amount of angular momentum outward (Manikantan et al., in prep). This highlights the importance of magnetic flux eruptions in global disk evolution. 

Even though we have limited ourselves to a particular type of initial conditions, i.e., a geometrically-thick hydrodynamic torus with poloidal magnetic fields around a non-spinning BH, our results are applicable for sub-Eddington accreting MAD flows in general, be it when the inflow is nearly spherical, or stellar wind-fed or even geometrically thin. It will be interesting to check how magnetic flux-tubes interact with the infalling gas for these different accretion modes, especially for slowly rotating inflows, since flux-tubes may be able to travel further out to larger radii before they undergo dissipation due to shearing. Given that the theoretical model comparison to both the near-horizon structure of the supermassive BHs M87$^*$ and Sgr A$^*$ \citep[][]{EHT_M87_2019_PaperV, EHT_SgrA_2022_PaperV} favored a magnetically dominated inflow, it is highly probable that magnetic flux eruptions regulate both mass and angular momentum loss in these systems, apart from being a potential mechanism behind the production of high energy flares \citep[e.g.,][]{Dexter:2020:NIR_MAD,Porth:2020:NIR_MAD,Ripperda2022,Scepi:2022}.

\section*{Acknowledgements}
\label{sec:acks}
\noindent We thank the anonymous referee for their thoughtful comments and suggestions. This research was enabled by support provided by grant no. NSF PHY-1125915 along with a INCITE program award PHY129, using resources from the Oak Ridge Leadership Computing Facility, Summit, which is a DOE office of Science User Facility supported under contract DE-AC05- 00OR22725, and Calcul Quebec (http://www.calculquebec.ca) and Compute Canada (http://www.computecanada.ca). KC and RN are supported by the Black Hole Initiative at Harvard University, which is funded by grants from the Gordon and Betty Moore Foundation, John Templeton Foundation and the Black Hole PIRE program (NSF grant OISE-1743747). The opinions expressed in this publication are those of the authors and do not necessarily reflect the views of the Moore or Templeton Foundations. This research has made use of NASA’s Astrophysics Data System.



\bibliographystyle{aasjournal}
\bibliography{newbib}
\end{document}